\documentclass[article]{jss}

\usepackage{orcidlink,thumbpdf,lmodern}

\usepackage{framed}
\usepackage{booktabs}
\usepackage{multirow}

\usepackage{amsmath, amsfonts, amssymb}
\usepackage{tabularx}
\usepackage{pifont}
\usepackage[all]{nowidow}  % avoid widows and orphans
\usepackage[normalem]{ulem}  % extra underlining

\usepackage[dvipsnames]{xcolor}
\newcommand{\data}{\mathcal{D}}
\newcommand{\model}{M}
\newcommand{\argmin}{\operatorname*{argmin}}
\newcommand{\obs}{\operatorname{obs}}
\newcommand{\yes}{\color{ForestGreen}\ding{51}}
\newcommand{\no}{\color{WildStrawberry}\ding{55}}
\newcommand{\yesnt}{\color{ForestGreen}(\ding{51})}

\usepackage{graphicx}

\newcommand{\py}{\raisebox{-0.2\height}{%
  \includegraphics[
    height=1.6ex,
    width=2.2ex,
    keepaspectratio
  ]{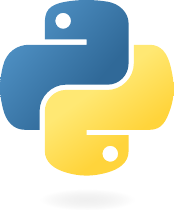}%
}}

\newcommand{\jl}{\raisebox{-0.2\height}{%
  \includegraphics[
    height=1.6ex,
    width=2.2ex,
    keepaspectratio
  ]{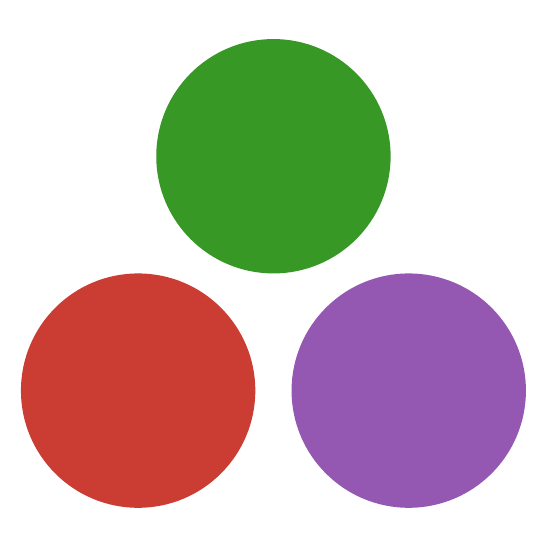}%
}}

\author{
    Lars Kühmichel~\orcidlink{0000-0001-7771-3525}\\
    TU Dortmund University\\
    \And
    Hans Olischläger~\orcidlink{0009-0002-1285-2959}\\
    TU Dortmund University\\
    \AND
    Jonas Arruda~\orcidlink{0009-0008-9644-5771}\\
    University of Bonn\\
    \And
    Jerry M. Huang~\orcidlink{0009-0004-8391-6141}\\
    Rensselaer Polytechnic Institute\\
    \And
    Daniel Habermann~\orcidlink{0000-0003-3685-7287}\\
    TU Dortmund University\\
    \AND
    Valentin Pratz~\orcidlink{0000-0001-8371-3417}\\
    Heidelberg University\\
    \And
    \v{S}imon Kucharsk\'{y}~\orcidlink{0000-0003-4192-1140}\\
    TU Dortmund University\\
    \And
    Lasse Elsemüller~\orcidlink{0000-0003-0368-720X}\\
    Independent Scientist\\
    \AND
    Aayush Mishra~\orcidlink{0000-0003-1164-0268}\\
    TU Dortmund University\\
    \And
    Niels Bracher~\orcidlink{0009-0001-9680-0875}\\
    Rensselaer Polytechnic Institute\\
    \And
    Svenja Jedhoff~\orcidlink{0009-0008-2939-9103}\\
    TU Dortmund University\\
    \AND
    Marvin Schmitt~\orcidlink{0000-0002-7967-4723}\\
    Independent Scientist\\
    \And
    Paul-Christian Bürkner~\orcidlink{0000-0001-5765-8995}\\
    TU Dortmund University\\
    \And
    Stefan T. Radev~\orcidlink{0000-0002-6702-9559}\\
    Rensselaer Polytechnic Institute\\
}
\Plainauthor{Lars Kühmichel et al.}

\title{\vspace{-1cm}\pkg{BayesFlow} 2: Multi-Backend Amortized Bayesian Inference in \proglang{Python}}
\Plaintitle{BayesFlow 2: Multi-Backend Amortized Bayesian Inference in Python}
\Shorttitle{BayesFlow 2}
\Abstract{
Modern Bayesian inference involves a mixture of computational methods for estimating, validating, and drawing conclusions from probabilistic models as part of principled workflows. An overarching motif of many Bayesian methods is that they are relatively slow, which often becomes prohibitive when fitting complex models to large data sets. Amortized Bayesian inference (ABI) offers a path to solving the computational challenges of Bayes.
ABI trains neural networks on model simulations, rewarding users with rapid inference of any model-implied quantity, such as point estimates, likelihoods, or full posterior distributions.
In this work, we present the \proglang{Python} library \pkg{BayesFlow}, Version 2, for general-purpose ABI. 
Along with direct posterior, likelihood, and ratio estimation, the software includes support for multiple popular deep learning backends, a rich collection of generative networks for sampling and density estimation, complete customization and high-level interfaces, as well as new capabilities for hyperparameter optimization, design optimization, and multilevel modeling. Using a case study on dynamical system parameter estimation, combined with comparisons to similar software, we show that our streamlined, user-friendly workflow has strong potential to support broad adoption.
}

\Plainkeywords{Amortized Bayesian Inference, Simulation-Based Inference, Deep Learning}
\Address{
    Lars Kühmichel\\
    Department of Statistics\\
    TU Dortmund University\\
    44227 Dortmund, Germany\\
    E-mail: \email{lars.kuehmichel@tu-dortmund.de}\\
    URL: \url{https://github.com/LarsKue}
}

\makeatletter
\def\jssfix{\global\let\jss@And\And \global\let\jss@AND\AND
  \gdef\And{\noalign{\vskip-10pt}\jss@And}\gdef\AND{\noalign{\vskip-10pt}\jss@AND}}
\expandafter\def\expandafter\@author\expandafter{\expandafter\jssfix\@author}
\makeatother
\begin{document}

\newpage

\section{Introduction}

Simulation-based inference \citep[SBI;][]{cranmer2020frontier, Deistler2025-sbitutorial} has become an increasingly prevalent element in computational science \citep{lavin2021simulation}. 
SBI addresses the problem of estimating unknown parameters $\theta$ from data $\mathcal{D}$ based on a probabilistic model $p(\theta, \mathcal{D})$ that can be simulated.
Although it was originally motivated in settings where traditional Bayesian methods are intractable or infeasible \citep{diggle1984monte}, SBI is steadily emerging as a general approach to Bayesian computation \citep{Zammit-Mangion2025-review}.

In many instances, we may want to fit a model to multiple data sets, be it to analyze big data \citep{vonkrause2022mentalspeed} or to verify key properties of the model (e.g., identifiability) \textit{in silico} \citep{burkner2025simulations}.
Thus, it becomes desirable to pool or ``compile'' computations into global estimators that can produce near-instant results for arbitrary queries, unless we dispose of infinite time of computational resources.
As a general framework for efficient probabilistic reasoning and model inversion \citep{gershman2014amortized, stuhlmuller2013learning, kingma2013auto, paige2016inference, le2017inference},
\textit{amortized inference} achieves this by learning estimators to answer many queries with minimal recomputation.

As the emerging default choice in SBI workflows, amortized Bayesian inference (ABI) trains neural networks on simulations from $p(\theta, \mathcal{D})$. The cost of this training \textit{amortizes} by reusing the networks for rapid inference of any model-implied quantity, such as point estimates, likelihoods, or full posterior distributions.
As such, ABI offers both alternative and complementary solutions to established inference methods, such as Markov chain Monte Carlo \citep[MCMC;][]{brooks2011handbook} and non-amortized variational inference \citep{ranganathBlackBoxVariational2014}, or Laplace approximation \citep{rue2009approximate}, especially when likelihood evaluations are computationally expensive or intractable.

Version 2 of \pkg{BayesFlow} represents a complete redesign that substantially extends and improves upon the initial release \citep{radev2023bayesflow}.
It adopts the widespread shift from isolated analyses to iterative Bayesian workflows \citep{gelman2020bayesian, li2024amortized}.
Further, it aligns with the growing interest in AI-assisted statistics \citep{musslick2025automating}, where neural surrogates can replace any performance-critical component of classical workflows \citep{ye2025integrating}.

With a streamlined high-level user interface for ABI workflows, \pkg{BayesFlow} addresses the demand for accessible and flexible software solutions.
As part of the \pkg{Keras 3} ecosystem \citep{chollet2025keras3}, \pkg{BayesFlow} supports all popular deep-learning backends and integrates smoothly into modern ML workflows within \proglang{Python}.
It is highly portable and can also be accessed from \pkg{R} \citep{R2025}, enhancing interoperability with the \pkg{R} ecosystem.
With robust default settings, it lowers the entry barrier for new users.
Most importantly, \pkg{BayesFlow} covers the full Bayesian workflow—from building and fitting complex models (e.g., likelihood-free or likelihood-based, multilevel or flat, dynamic or stationary, \textit{etc}) to model comparison, criticism, and deployment.

\section{Amortized Bayesian Inference}
\label{sec:abi}

\begin{figure}[t]
    \centering
    \includegraphics[width=0.99\linewidth]{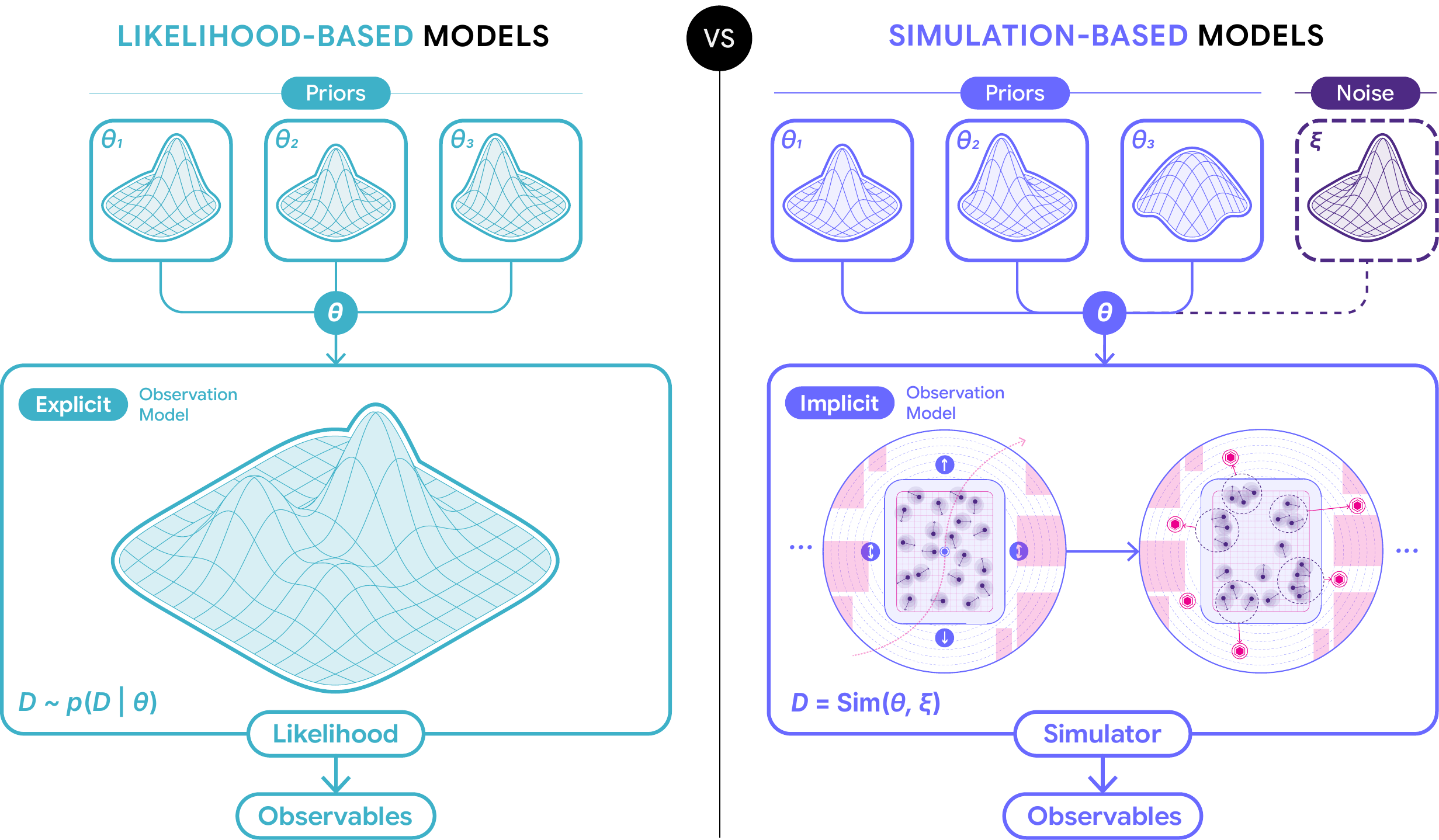}
    \caption{Side-by-side comparison of likelihood-based (\textit{explicit}) and simulation-based (\textit{implicit}) models from a Bayesian perspective. In likelihood-based models, the data model $p(\data \mid \theta)$ and prior $p(\theta)$ can be explicitly sampled and evaluated. In simulation-based models, the data model $p_{\text{implicit}}(\data \mid \theta)$ and prior $p(\theta)$ can be sampled through stochastic simulation, but cannot be (easily) evaluated.}
    \label{fig:abi}
\end{figure}

\subsection{Learning From Simulations}

BayesFlow expects the modeler to specify a probabilistic model $p(\theta, \mathcal{D})$ over empirical (observable) quantities $\mathcal{D}$ and latent (unobservable) quantities $\theta$.
Traditionally, the joint model is completely defined by a \textit{likelihood function} $p(\data \mid \theta)$ and a \textit{prior} $p(\theta)$:
\begin{equation}
    p(\theta, \mathcal{D}) = p(\data \mid \theta) p(\theta).
\end{equation}
In contrast to modern MCMC-based software such as \proglang{Stan} \citep{stan_2025} or \pkg{PyMC} \citep{abril_pymc_2023}, \pkg{BayesFlow} does not need to evaluate any of these distributions for inference. Instead, it only relies on the ability to \textit{simulate} from the joint model $p(\theta, \mathcal{D})$ according to:
\begin{equation}
    \mathcal{D} = \text{Sim}(\theta, \xi), \quad \theta \sim p(\theta), \; \xi \sim \text{RNG}(\cdot) \quad \Longleftrightarrow \quad \mathcal{D} \sim p_{\text{implicit}}(\mathcal{D} \mid \theta), \quad \theta \sim p(\theta),
\end{equation}
where $\xi \in \Xi$ denotes random (outsourced) noise variates and $p_{\text{implicit}}(\mathcal{D} \mid \theta)$ denotes the likelihood \textit{implied} by the left-hand side. Under certain conditions, this implicit likelihood can be written as an integral over the stochastic components of the simulator:
\begin{equation}
    p_{\text{implicit}}(\mathcal{D} \mid \theta) = \int_{\Xi} p(\mathcal{D}, \xi \mid \theta) \, \text{d}\xi,
\end{equation}
which is typically intractable and motivates the application of simulation-based inference \citep[SBI;][]{cranmer2020frontier}.
In modern SBI, simulations serve as training data for specialized neural networks. Once trained, these networks can infer any model quantity of interest (e.g., parameters) with minimal overhead, amortizing the initial investment in training the networks. This approach applies to any model from which joint samples $(\theta, \mathcal{D})$ can be drawn, regardless of whether the likelihood is available in closed form, is evaluable only up to a constant, or is intractable (see~\autoref{fig:abi}); the only requirement is a parameterized simulator.

% \begin{table}[t!]
% \centering
% \label{tab:likelihood_vs_simulator}
% \resizebox{\textwidth}{!}{%
% \begin{tabular}{@{}lp{0.45\textwidth}p{0.45\textwidth}@{}}
% \toprule
%  & \textbf{Likelihood-Based (Explicit)} & \textbf{Simulation-Based (Implicit)} \\
% \midrule
% \textbf{Model specification} &
% \(\theta \sim p(\theta),\quad \data \sim p(\data \mid \theta)\) &
% $\theta \sim p(\theta),\quad \data = \mathrm{Sim}(\theta, \xi), \, \xi \sim \text{RNG}(\cdot)$ \\
% \addlinespace
% \textbf{Prior} $p(\theta)$ &
% can be sampled and evaluated &
% can be sampled and \emph{optionally} evaluated \\
% \addlinespace
% \textbf{Data model} $p(\data \mid \theta)$ &
% can be sampled and evaluated &
% can be sampled but \emph{not} evaluated \\
% \addlinespace
% \textbf{Posterior} $p(\theta \mid \data)$ & usually intractable &
% doubly intractable \\
% \bottomrule
% \end{tabular}
% }
% \caption{Side-by-side view of likelihood-based (\textit{explicit}) and simulation-based (\textit{implicit}) models from a Bayesian perspective. \pkg{BayesFlow} is applicable to both model classes.}
% \end{table}

\subsection{Bayesian Parameter Estimation}

BayesFlow adopts a Bayesian framework for parameter estimation, enabling inference over the full posterior distribution:
\begin{equation}\label{eq:post}
    p(\theta \mid \data) \propto p(\data \mid \theta) p(\theta).
\end{equation}
We cast posterior learning as minimizing an expected objective over joint draws $(\theta, \data) \sim p(\theta, \data)$:
\begin{align}
    \hat{q} &= \argmin_q \mathbb{E}_{(\theta, \data) \, \sim \, p(\theta, \data)} \left[\mathcal{J} \big(q(\cdot\mid \data),\,\theta \big) \right] \label{eq:expected_score_general} \\
    &\approx \argmin_q \frac{1}{B}\sum_{b=1}^B \mathcal{J}(q(\cdot \mid \data^{(b)}), \theta^{(b)}),
\end{align}
where $B$ is the \textit{simulation budget}, $\mathcal{J}$ is a \textit{strictly proper scoring rule} (e.g., the log score), and $q$ is the approximate distribution defined by a generative neural network.
Different choices of the objective $\mathcal{J}$ recover well-known estimators.
When $q$ is a conditional density estimator trained with this objective, the approach is known as \textit{neural posterior estimation} \citep[NPE;][]{papamakarios2016fast, greenberg2019automatic, radev2020bayesflow} and constitutes the most common use of \pkg{BayesFlow}.
Using the log score yields the maximum likelihood objective \citep{papamakarios2021normalizing}, whereas using a score-based objective corresponds to denoising score matching \citep{arruda2025diffusion}.
Although the latter does not rely on proper scoring rules, it can be interpreted as a surrogate objective that approximates maximum likelihood training under certain conditions \citep{song2021maximum}.

\pkg{BayesFlow} can also learn arbitrary Bayesian point estimators \citep{sainsbury2024likelihood}, such as posterior means, variances, or quantiles.
The resulting optimization objective simply replaces the approximate distribution $q$ in Eq.~\ref{eq:expected_score_general} with the desired point estimate $\hat{f}(\mathcal{D})$:
\begin{equation}
    \hat{f} = \argmin_f \mathbb{E}_{(\theta, \data) \, \sim \, p(\theta, \data)} \left[\mathcal{J} \big(f(\data),\,\theta \big) \right] \label{eq:point_score_general},
\end{equation}
where $\mathcal{J}$ is now a scoring rule for point estimators. In either case, inference is amortized: once trained, $\hat{q}$ can efficiently sample from the approximate posterior or $\hat{f}$ can produce the point estimates for any upcoming data $\data_{\rm new}$ compatible with the joint model $p(\theta, \data)$.

\subsection{Bayesian Model Verification}

Model verification refers to assessing a model’s assumptions \textit{in silico}, prior to applying it to real data where those assumptions may not hold \citep{burkner2025simulations}.
It typically involves estimating posteriors from simulated data and verifying key properties such as calibration (i.e., whether credible intervals are trustworthy) and recovery (i.e., whether parameters can be accurately estimated).
Traditionally associated with the high computational cost of model refitting, model verification in \pkg{BayesFlow} can be performed essentially for free due to amortization. Below, we discuss two key aspects of model verification.

\paragraph{Calibration} Consider a target quantity of interest $T = T(\theta)$ that is a function of the parameters $\theta$. This could include the parameters themselves or any pushforward quantity, such as posterior predictions. If the model is well-specified (i.e., is equal to the true data-generating distribution), then the following equality holds:
\begin{equation}
\label{prior-cond-sbc}
  \alpha = \mathbb{E}_{(\tilde{\theta}, \data) \, \sim \, p(\theta, \data) } \left[ \mathbb{I}(T(\tilde{\theta}) \in U_\alpha(T(\theta) \mid \data)) \right],
\end{equation}
for all target quantities $T$ and all uncertainty regions $U_\alpha(T(\theta) \mid \data)$ obtained from the analytic posterior $p(\theta \mid \data)$ with nominal coverage probability $\alpha$,
where $\tilde{\theta}$ denotes a simulated ground truth \citep{burknerModelsAreUseful2023}.
In other words, the probability that an uncertainty region (e.g., a credible interval) with coverage probability $\alpha$ contains the true value must be equal to $\alpha$ on average. This property can be used to verify the calibration of an approximate posterior. If the approximation is good enough, then Eq.~\eqref{prior-cond-sbc} should hold for the uncertainty regions implied by the approximation \citep{cook2006validation, talts2020sbc}. 

Such checks provide necessary but not sufficient conditions---a failed check indicates miscalibration, whereas a passed check does not by itself certify correctness \citep{modrakSimulationBasedCalibrationChecking2023}. However, verifying calibration requires fitting the model to many (usually hundreds of) simulated data sets \citep{modrakSimulationBasedCalibrationChecking2023}. This can be computationally challenging for MCMC-based or other non-amortized algorithms. In \pkg{BayesFlow}, amortization provides a decisive advantage: once trained, a posterior estimator can self-diagnose efficiently across the entire sample space, making simulation-based calibration (SBC) routine.

\paragraph{Parameter recoverability}
Another key aspect of model verification concerns whether the model can accurately recover its own parameters. Formally, recoverability can be expressed as
\begin{equation} \text{Recovery}(\mathcal{L}) := \mathbb{E}_{(\tilde{\theta}, \data) \, \sim \, p(\theta, \data)} \left[\int_{\Theta} \mathcal{L}(\tilde{\theta}, \theta) \, p(\theta \mid \data)\,\text{d}\theta \right], 
\end{equation}
where $\mathcal{L}$ denotes a loss metric (e.g., root mean square error; RMSE). Related summaries of the posterior relative to the prior, such as the posterior $z$-score and posterior contraction, are discussed by \cite{betancourt2018calibrating} and \cite{schad2021toward}.
Analogous to calibration, traditional recovery analyses are computationally demanding due to repeated posterior inference over many simulated datasets \citep{schad2021toward}. In contrast, \pkg{BayesFlow} can analyze recovery essentially instantaneously after training.
%amortization again provides a decisive advantage: once trained, a posterior or point estimator can evaluate recoverability metrics efficiently across the entire sample space, making comprehensive model verification routine.

\subsection{Likelihood Estimation}

In some applications, learning a surrogate $q(\data \mid \theta)$ for the intractable likelihood $p_{\text{implicit}}(\data \mid \theta)$ may be desirable. Learning the likelihood decouples model training from prior specification and thus enables flexible reuse, for instance, in multilevel models or regression models where parameters $\theta$ are functions of other variables \citep{papamakarios2019sequential, lueckmann2019likelihood, fengler2021likelihood, boelts2022flexible}. Additionally, a neural likelihood can serve as a fast, differentiable \textit{emulator} for an otherwise expensive simulator.

From the perspective of \pkg{BayesFlow}, neural likelihood estimation merely amounts to swapping the arguments in Eq.~\ref{eq:expected_score_general}, where now the approximate density $q(\cdot \mid \theta)$ is conditioned on the parameters. Learning the likelihood may require larger simulation budgets if the dimensionality of $\data$ is high relative to that of $\theta$.
Moreover, if the goal is to eventually estimate posteriors, the resulting pipeline is \textit{not fully} amortized, since it requires downstream MCMC for every $\data$ \citep{radev2023jana}. 

Related to likelihood estimation is likelihood-to-evidence ratio estimation, approximating the density ratio $p(\data \mid \theta) \, / \, p(\data)$ \citep{hermansLikelihoodfreeMCMCAmortized2020, durkanContrastiveLearningLikelihoodfree2020, millerContrastiveNeuralRatio2022}. Since this ratio is proportional to the likelihood, a likelihood-to-evidence ratio estimate can be used in place of the likelihood density, for example, for subsequent posterior estimation with MCMC. \pkg{BayesFlow} implements the stable contrastive training of \cite{millerContrastiveNeuralRatio2022}, which uses a simple multiclass classifier.

\subsection{Bayesian Model Comparison}

The preceding probabilistic quantities depend implicitly on various model assumptions. These can be abstractly denoted as $\model$, and comparing the utility of different assumptions requires Bayesian model comparison \citep[BMC;][]{mackay2003information}.
Formally, \textit{prior predictive} BMC depends on a model's marginal likelihood (i.e., the normalizer of Eq.~\ref{eq:post}):
\begin{equation}
    p(\data \mid \model) = \int_{\Theta} p_{\text{implicit}}(\data \mid \theta, \model)\,p(\theta \mid \model)\,\text{d}\theta,
\end{equation}
which is doubly intractable for implicit models. Since marginal likelihoods are normalized probability distributions, they penalize model complexity by trading off sharpness for dispersion: complex models can generate a broader range of data, but must distribute their probability mass across a greater volume of the sample space.

The ratio of marginal likelihoods for two models is known as the Bayes factor \citep[BF;][]{kass1995bayes}:
\begin{equation}
\text{BF}_{12} := \frac{p(\data \mid \model_1)}{p(\data \mid \model_2)} = \frac{p(\model_1 \mid \data)}{p(\model_2 \mid \data)} \; \Big/ \; \frac{p(\model_1)}{p(\model_2)}.
\end{equation}
The first term on the right expresses the posterior odds, while the second term represents the prior odds.
%When comparing $J$ models, we need to approximate $J(J - 1) / 2$ Bayes factors, which are typically presented as a Bayes factor matrix. 
For a set of $J$ models, we can equivalently compute the corresponding $J$ posterior model probabilities $p(\model_j \mid \data)$ assuming that the true model is within the set of considered models.
\pkg{BayesFlow} estimates the  posterior model probabilities by treating BMC as a probabilistic classification task:
\begin{equation}
\hat{q} = \argmin_q \mathbb{E}_{(M, \theta, \data) \, \sim \, p(M)p(\theta, \data\,\mid\,M)} \left[\mathcal{J} \big(q(\cdot \mid \data), M \big) \right],
\end{equation}
where $\mathcal{J}$ denotes a proper scoring rule (e.g., loss or log score) and $M$ is the true model index. In essence, we define a supervised learning problem over simulated pairs $(M, \data)$ and train a probabilistic classifier to predict model labels \citep{radev2021evidential, pudlo2016reliable, jeffrey2024evidence}.

Alternatively, \pkg{BayesFlow} can approximate the log marginal likelihood (LML) using a pair of trained posterior and likelihood estimators and rearranging Bayes' rule:
\begin{equation}\label{eq:lml}
    \log \hat{q}(\data \mid \model) = \log \hat{q}(\data \mid \theta, \model) + \log p(\theta \mid \model) - \log \hat{q}(\theta \mid \data, \model).
\end{equation}
Since the quantity on the left-hand side is constant with respect to $\theta$, we can theoretically plug in every $\theta \in \Theta$ in the right-hand side and obtain an estimate of the LML. However, due to imperfect posterior and likelihood estimation, different values of $\theta$ will lead to different values of $\log \hat{q}(\data \mid \model)$, providing a measure of approximation error in addition to a point estimate of the LML. This error can be reduced by using the analytic likelihood whenever available \citep{radev2023jana, kucharskyImprovingAccuracyAmortized2025}.

Finally, we can also rank models according to \textit{posterior predictive} quantities computed from new data $\tilde{\data}$, such as the expected log predictive density \citep[ELPD;][]{vehtariSurveyBayesianPredictive2012}:
\begin{equation}
    \text{ELPD}(M) = \mathbb{E}_{\tilde{D} \, \sim \, p(\data)} \left[ \log \int_{\Theta} p_{\text{implicit}}(\tilde{\data} \mid \theta, M)\,p(\theta \mid \data)\text{d}\theta \right]. 
\end{equation}
The ELPD is triply intractable, but the outer expectation can be estimated using cross-validation, whereas the inner expectation can be estimated by averaging a likelihood surrogate over approximate posterior samples \citep{radev2023jana}.

\subsection{Bayesian Sensitivity Analysis}

The results of any (Bayesian) analysis are sensitive to modeling choices related to likelihood specification, prior elicitation, and data processing. For instance, a common question in Bayesian analysis is: How would the results look under a different prior? Sensitivity analysis provides a formal answer to such questions and typically requires refitting the model under different assumptions. The resulting computational bottleneck can be resolved in \pkg{BayesFlow} by extending the \textit{amortization scope} with context variables $C$ that indicate potential factors of inferential variation. For example, $C$ can encode the width of the prior or represent an entirely different model specification.

To enable amortized, sensitivity-aware analysis, we can extend the joint model to $p(C, \theta, \data)$ and make $C$ an additional input to the neural estimator. Sensitivity-aware estimation then simply extends the objective in Eq.~\eqref{eq:expected_score_general} to include $C$:
\begin{equation}
\hat{q} = \argmin_q \mathbb{E}_{(C, \theta, \data) \, \sim \, p(C, \theta, \data)} \left[\mathcal{J} \big(q(\cdot\mid \data, C), \theta \big) \right].
\end{equation}
During inference, the modeler can supply different values of $C$ and assess the resulting differences qualitatively or quantitatively \citep{elsemuller2024sensitivityaware}. Additionally, in order to estimate the sensitivity of results with regard to network and training hyperparameters, one can train an ensemble of networks and subsequently analyze the variation across ensemble members.

\subsection{Handling Model Misspecification}

In countless settings, researchers model complex phenomena with misspecified models \citep{walkerBayesianInferenceMisspecified2013}.
There are many criteria for determining whether a misspecified (read: wrong) model is nevertheless a useful one \citep{burknerModelsAreUseful2023}.
In the context of SBI, model misspecification translates into simulated data that lacks relevant structure present in the real data.
Simulation gaps are not just evidence that the model is misspecified; when they are severe, the trained network may no longer estimate the same quantity it was designed to estimate during training \citep{schmitt2023detecting, gloeckler2023adversarial, frazier2024statistical}.

A useful way to see this as a unique problem is to contrast amortized inference with MCMC. If an MCMC sampler mixes and converges, it will (by construction) return the posterior implied by Bayes’ rule for the assumed model, even if that posterior is a poor representation of reality. Amortized neural estimators are different: under a simulation gap, they can drift away from the Bayes posterior of the nominal model and instead output something that reflects the training distribution and the network’s inductive biases. In that case, the posterior estimate $q(\theta \mid \mathcal{D})$ may no longer correspond to any coherent Bayesian update for the model, which undermines how we interpret the reported uncertainties and point estimates.

Consequently, detecting and mitigating model misspecification is now widely viewed as an integral part of an amortized Bayesian workflow \citep{li2024amortized}. \pkg{BayesFlow} provides tools to \textit{detect} potential simulation gaps by inspecting the typicality of real data relative to the distribution of learned data embeddings or summary statistics $\phi(\mathcal{D})$. This can be done using any out-of-distribution (OOD) or outlier detection score as a proxy measure of potential ``extrapolation bias'' \citep{frazier2024statistical}.
The embedding distribution $p(\phi(\mathcal{D}))$ can be further ``Gaussianized'' via an additional training loss \citep{schmitt2023detecting} to meet the assumptions of simple parametric methods, such as the Mahalanobis distance \citep{li2024amortized}.

Mitigating the impact of model misspecification in amortized estimators remains an active field of research. Sometimes, the simplest solution is to assume a corrupted model $p_{\epsilon}(\theta, \mathcal{D})$ that can simulate outliers and induce a robust neural estimator \citep{ward2022robust}, trading off some accuracy for much higher breakdown points \citep{wu2024testing}. Additional loss functions, such as Bayesian self-consistency \citep{mishra2025robust} or unsupervised domain adaptation \citep{huang_learning_2023, elsemueller2025does} can improve robustness as well.

Alternatively, for datasets flagged as OOD, one can try to correct the posteriors \textit{post hoc} using Pareto-smoothed importance sampling (PSIS) when a likelihood is available \citep{vehtari_pareto_2024}, or more generally via inference-time adaptation \citep{siahkoohi2023reliable}.
Crucially, \pkg{BayesFlow} implements all components of the amortized Bayesian workflow \citep{li2024amortized}, along with spike-and-slab corruption \citep{ward2022robust}, latent space maximum mean discrepancy \citep[MMD;][]{schmitt2023detecting}, and self-consistency losses \citep{schmitt_leveraging_2024, mishra2025robust}.

\section{Software}
\label{sec:software}

\begin{figure}[t]
    \centering
    \includegraphics[width=0.99\linewidth]{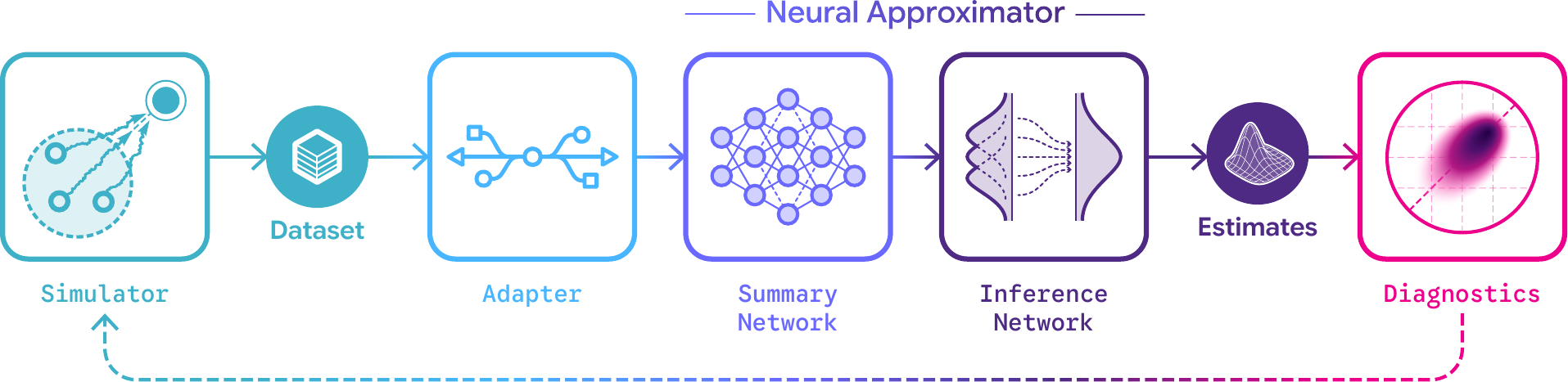}
    \caption{Overview of the basic amortized Bayesian workflow provided by \pkg{BayesFlow}. The \code{Simulator} interface provides users with automated grouping of data-generating functions and simulated data. The \code{Adapter} interface transforms the raw simulator outputs for training the neural estimator. The neural estimator typically consists of a \code{SummaryNetwork} that encodes observables into a latent summary vector, and an \code{InferenceNetwork} conditioned on this vector to approximate target distributions. The resulting estimates are then used for model validation through \pkg{BayesFlow}'s \code{diagnostics} module, with a wide array of graphical and numerical diagnostics for checking computational faithfulness and model sensitivity. The entire workflow is encapsulated at the high level via the \code{Workflow} object, which allows users to rapidly iterate through the Bayesian workflow.}
    \label{fig:workflow}
\end{figure}

The \pkg{BayesFlow} software is divided into modules, which together enable complete end-to-end ABI workflows. Below, we introduce all the main modules of \pkg{BayesFlow} (see also~\autoref{fig:workflow}). 

\subsection{Simulators}
\label{ssec:simulators}

Since ABI performs simulation-based inference, simulators play a critical role in \pkg{BayesFlow}. The package offers both lower and higher level simulator interfaces that provide users with different options for balancing expressivity, computational efficiency, and ease-of-use.
%With the above interfaces, in principle arbitrary simulators can be expressed. While the higher-level (auto-batched) interface is sufficient for most cases of low to medium complexity, the lower-level interface offers full control over all details of the simulation process if required.

On a lower-level, simulators in \pkg{BayesFlow} are specified as a class with a \code{sample(batch_size)} method, returning a dictionary of data in the form of \pkg{NumPy} arrays. The keys of the dictionary are arbitrary, user-chosen strings, denoting the variable names. For example, consider the simple probabilistic model given by
\begin{align*}
    \mu &\sim \text{Normal}(0, 1) \\
    \sigma &\sim \text{Exponential}(1) \\
    x_n &= \text{Normal}(\mu, \sigma) \quad n = 1,\ldots,N
\end{align*}

This can be coded in the form of a \pkg{BayesFlow} simulator as

\begin{Code}
import numpy as np
import bayesflow as bf

class MySimulator(bf.simulators.Simulator):
    def sample(self, batch_size, N=10):
        mu = np.random.normal(size=(batch_size, 1))
        sigma = np.random.exponential(size=(batch_size, 1))
        x = np.random.normal(size=(batch_size, N), loc=mu, scale=sigma)
        return {"mu": mu, "sigma": sigma, "x": x}
\end{Code}

We can then simulate (a batch of) parameter-data pairs from the simulator via

\begin{Code}
simulator = MySimulator()
sims = simulator.sample(batch_size=5)
\end{Code}

For computational efficiency, \pkg{BayesFlow} assumes that a single call to \code{sample(batch_size)} will return a batched dictionary of data, that is, the shape of all value arrays is \code{(batch_size, ...)}. For the above example, \code{sims["mu"]} and \code{sims["sigma"]} have shape \code{(batch_size, 1)} and \code{sims["x"]} has shape \code{(batch_size, N)}. 
%The need for a \code{batch_size} arises to enable mini-batch optimization, which essential for most deep learning applications.
The need for a \code{batch_size} may not be intuitive for users without a deep learning background, but is required for mini-batch optimization, essential for most deep learning applications.
%since it requires them to simulate not only one but a whole batch of parameter-data pairs in one go. %it can sometimes be cumbersome to implement simulators in a batched fashion. 

For convenience, \pkg{BayesFlow} also supplies users with a utility to auto-batch simulators starting from unbatched functions: \code{bf.make_simulator}. Using this higher-level simulator interface, the same probabilistic model as above can now be coded as

\begin{Code}
def prior():
    mu = np.random.normal()
    sigma = np.random.exponential()
    return {"mu": mu, "sigma": sigma}

def likelihood(mu, sigma, N=10):
    x = np.random.normal(size=N, loc=mu, scale=sigma)
    return {"x": x}

simulator = bf.make_simulator([prior, likelihood])
\end{Code}

In the above simulator, the sample size \code{N} is considered fixed for the whole batch. This is important as \code{N} influences the shape of \code{x} and simulations can only be safely concatenated into batches if they have the same shape for each simulator run. In order to also vary \code{N} within the simulator, we need to vary it \textit{by batch} and not by simulation within batch. For this purpose, \code{bf.make_simulator} provides the optional \code{meta_fn} argument, which takes a function that returns a dictionary of variables that should only vary by batch (and be constant within batches). 
%This workflow also supports batch-level metadata via an optional \code{meta_fn} function. This is a callable that returns a dictionary of data that is constant for the whole batch. 
If provided, \pkg{BayesFlow} calls \code{meta_fn} once per batch and prepends its outputs to the regular simulator call. Extending our above simulator to sample $N$ uniformly between $10$ and $20$, we can write 
%This is useful when the generative process depends on context such as varying observation counts, e.g.:

\begin{Code}
def meta():
    return {"N": np.random.randint(10, 21)}

simulator = bf.make_simulator([prior, likelihood], meta_fn=meta)
\end{Code}

The output of \code{simulator.sample()} now contains a new entry \code{"N"} whose value is just a single integer: the sample size used for all simulations within the batch. To increase speed, simulations can also be run in parallel with the \code{simulator.sample_parallel()} method.

For models with more complex probabilistic factorizations than just a single set of parameters implying data, one can use the
\code{HierarchicalSimulator} class, among others, enabling simulations for multilevel and mixture models \citep{habermann_amortized_2024, habermann2026multilevel, kucharsky2025amortized}.

\subsection{Adapters}
\label{ssec:adapters}

The raw simulator outputs are not immediately usable to train neural networks. Rather, multiple pre-processing steps have to be performed first. This specifically includes (a) bringing variables into correct shapes (e.g., broadcasting, splitting, or concatenating variables), (b) constraining variables (e.g., enforcing non-negativity), and (c) informing the networks which variables are to be used for which purpose (e.g., which variables are parameters and which are data). \pkg{BayesFlow} provides a single interface for all these pre-processing tasks: the \code{Adapter}, which takes a dictionary as input (usually produced by a \code{Simulator}) and subsequently returns a dictionary of transformed variables.

To make this more concrete, for our simple model above, we need to (1) broadcast \code{"N"} to the shape of \code{"x"} since we eventually want to combine both as neural network conditions, (2) indicate that \code{"x"} is a set of exchangeable values whose order does not matter for inference, (3) constrain \code{"sigma"} to always be predicted as positive, and (4) square-root transform \code{"N"} for improved numerical stability. 

Additionally, we need to indicate which variables will have which role in the amortized inference task. In our example, \code{"mu"} and \code{"sigma"} are the parameters of interest, whose posterior we will eventually seek to infer conditional on data. As such, these two variables will be concatenated as \code{"inference_variables"}, a protected variable name understood by subsequent modules. 
The data \code{"x"} shall be first summarized by a summary network, which we indicate by renaming \code{"x"}
to \code{"summary_variables"} in the adapter. Lastly, we also want to condition our inference on the sample size \code{"N"}, but intend to pass it to the inference network directly as no summarization is needed, achieved by renaming \code{"N"} to \code{"inference_conditions"}. 
%An overview of protected variable names and their meanings is provided in Figure X. 
Putting all of these transforms together in the form of a single adapter is achieved by the following code:

%- Raw simulator output is not usable yet for inference
%- BayesFlow needs to be informed which variables should be inferred and which conditioned on (data flow - which variable goes into which network?)
%- Pandas-esque data pre-processing (e.g. constraining to sub-spaces, dropping keys, broadcasting, splitting or concatenating variables, ...)
%- BayesFlow provides a high-level interface that bridges the gap between simulator and estimator: Adapter

\begin{Code}
adapter = (
    bf.Adapter()
    .broadcast("N", to="x")
    .as_set("x")
    .constrain("sigma", lower=0)
    .sqrt("N")
    .concatenate(["mu", "sigma"], into="inference_variables")
    .rename("x", "summary_variables")
    .rename("N", "inference_conditions")
)
\end{Code}

In each of the above lines, an additional transform is added to the adapter, to be executed sequentially once fed with data. The shown syntax avoids having to repeat \code{adapter.<trans>} at every line. In more verbose form, the above code would read

\begin{Code}
adapter = bf.Adapter()
adapter = adapter.broadcast("N", to="x")
adapter = adapter.as_set("x")
...
\end{Code}

Once defined, we can directly pass the output of \code{simulator.sample} to \code{adapter}:

\begin{Code}
sims = simulator.sample(batch_size=5)
adapted_sims = adapter(sims)
\end{Code}

That said, users will rarely have to call adapters directly like this. Rather, adapters will be called automatically within the training and inference phase by other modules.

An important additional property of adapters is that most transforms are \textit{invertible}. That is, we could call

\begin{Code}
adapter(adapted_sims, inverse=True)
\end{Code}

to get back our original simulations. Some transforms are not invertible by definition and will thus be ignored during inversion. For example \code{adapter.drop("<variable name>")} drops a given variable from the dictionary, a transform that cannot be undone by inversion. In practice, invertibility is only required for transforms involving inference variables (here, parameters \code{"mu"} and \code{"sigma"}). This way, at inference time, the neural network predictions can traverse the adapter in inverse order, yielding the inferred variables with their original names and scales. 

%- Introduce the Adapter as the glue between Simulator and Estimator (makes the named-parameter data flow possible, by pseudo-invertibly mapping user-space parameters to network-space inputs)

%- Show how to define an Adapter using the high-level interfaces

%- Define the training strategy, and the data loading process (e.g., from disk vs. in-memory)

\begin{table}[htbp]
\centering
\begin{tabular}{@{}lllll@{}}
\toprule
\textbf{Generative Family} & \textbf{Architecture} & \textbf{Sampling} & \textbf{Density Evaluation} \\
\midrule
Normalizing Flows & Constrained & Single-step & Fast \\
Free-Form Flows & Semi-Constrained & Single-step & Moderate \\
Diffusion Models & Unconstrained & Multi-step & Slow \\
Flow Matching & Unconstrained & Multi-step & Slow \\
Consistency Models & Unconstrained & Few-step & N/A \\
\bottomrule
\end{tabular}
\caption{Generative neural network families available in BayesFlow.}
\label{tab:network_families}
\end{table}

\subsection{Networks}
\label{ssec:networks}
In most applications, \pkg{BayesFlow} employs two neural networks that together enable learning the target distribution: a \code{SummaryNetwork} that encodes (potentially variable-length) observations into a fixed-length latent summary vector, and an \code{InferenceNetwork} that conditions on this latent summary to produce an approximation of the target (see also~\autoref{fig:workflow}).

%Specifically, \pkg{BayesFlow} defines the \code{SummaryNetwork} as a simple feed-forward encoder that maps an exchangeable set or a time series of observations to a single latent embedding that encodes information from the whole set or series.

For a given application, the specific choice of \code{SummaryNetwork} should depend on the data that is supposed to be summarized. For example, when the observation set is exchangeable (as is $x$ in our simple model), common choices for the \code{SummaryNetwork} include pooling-based encoders \citep[e.g., \code{DeepSet};][]{deepset}, and attention-based set-encoders such as the \code{SetTransformer} \citep{settransformer}. These options trade computational cost for expressivity: pooling-based encoders are usually cheaper, whereas attention-based encoders may recover more complex inter-sample interactions at higher memory and compute cost. As another example, when the observations form a time-series, temporal neural networks, \code{TimeSeriesNetwork} \citep[LSTN;][]{zhang2023solving} or \code{TimeSeriesTransformer} \citep{wen2022transformers} are appropriate choices. A \code{wrappers} module allows for the incorporation of arbitrary summary networks written in a custom backend \citep[e.g., Mamba,][]{gu2024mamba}. Integrating foundation models as summary backbones \citep{pickens2026pre} is straightforward and forms the basis for one of the built-in tutorial.

The \code{InferenceNetwork} implements a conditional density estimator that (at minimum) allows sampling from the target distribution. Density evaluation is also supported, depending on the choice of architecture (see~\autoref{tab:network_families}). Usually, inference networks such as \code{FlowMatching} \citep{flowmatching,otflowmatching} are based on some sub-network architecture, such as a Multi-Layer Perceptrons (MLP), or a Convolutional Neural Network (CNN), that performs the actual feed-forward pass, while the wrapping \code{InferenceNetwork} implements the specific loss function, sampling process, and optionally density evaluation.

Appropriate choices for the \code{InferenceNetwork} depend strongly on the application. Usually, more expressive architectures, such as multi-step \code{DiffusionModel} \citep{diffusion, stablediffusion, song2020score, kingma2023understanding, arruda2025diffusion}, few-step \code{ConsistencyModel} \citep{song2023consistency, song2023improved, schmitt2024consistency} or \code{FlowMatching} trade sampling speed for accuracy. Single-step models such as \code{CouplingFlow}s \citep{nice,realnvp} allow for faster inference, but may struggle with dimensionality and multi-modality. In general, larger sub-networks can improve the expressivity of any \code{InferenceNetwork}, within the corresponding limits of the method.  Finally, free-form models (e.g., diffusion, flow matching) can implement arbitrary sub-nets that absorb the summary network inside their architectures.

To provide a concrete code example, we can define a pair of a \code{SetTransformer} summary network and a \code{CouplingFlow} inference network with depth 4 as follows:
\begin{Code}
summary_network = bf.networks.SetTransformer(summary_dim=8)
inference_network = bf.networks.CouplingFlow(depth=4)
\end{Code}
The \code{summary_dim} argument of \code{SummaryNetworks} is of particular importance since it defines the number of learned summaries to be extracted. As a rule of thumb, we recommend \code{summary_dim} to be at least twice or even four times as large as the combined dimensionality of the inference variables. For our example, we have $\mu$ and $\sigma$ as inference variables -- both of which are scalar, so their combined dimensionality is $2$. Choosing \code{summary_dim=8} should be a safe choice here.

\subsection{Estimators}
\label{ssec:estimators}

The \code{Estimator} class combines the target of inference, given by \code{Simulator} and \code{Adapter}, and the method of inference, given by the \code{Networks}. Within an \code{Estimator}, neural networks can be trained on the simulated data to subsequently perform inference on any new data. In most applications, all of our inference variables are continuous, so we will use the \code{BasicEstimator} subclass. There are also other estimator types, such as the \code{ScoringRuleEstimator} for learning arbitrary Bayes estimators (e.g., for posterior point estimation), \code{RatioEstimator} for learning likelihood-to-evidence ratios, \code{ModelComparisonEstimator} for learning posterior model probabilities (see \autoref{sec:abi}), and the experimental \code{MultilevelEstimator} (in \code{bayesflow.experimental}) for learning posteriors of models with complex probabilistic factorizations, such as multilevel and mixture models.

\subsubsection{Training of Estimators}
\label{sssec:training}

We initialize an estimator by supplying it with the previously defined \code{Networks} and the \code{Adapter}:

\begin{Code}
estimator = bf.estimators.BasicEstimator(
    inference_network=inference_network,
    summary_network=summary_network,
    adapter=adapter,
)
\end{Code}

After initialization, the \code{estimator} needs to be compiled with an appropriate optimizer. Since \pkg{BayesFlow} is built on \pkg{Keras3}, we use \code{keras} optimizers for this purpose. For example:

\begin{Code}
import keras

optimizer = keras.optimizers.AdamW(learning_rate=1e-4)
estimator.compile(optimizer=optimizer)
\end{Code}

After compilation, we are ready to train the neural networks via the \code{estimator.fit} method, which requires details on the training procedure, including the number of epochs (argument \code{epochs}), number of batches per epoch (\code{num_batches}) and the batch size (\code{batch_size}; number of simulations per batch). These are all standard arguments in deep learning pipelines.

Generally speaking, there are two types of training procedures: online training and offline training. Passing a \code{Simulator} to the \code{simulator} argument of \code{estimator.fit} implies online training, where simulated training data is generated on-the-fly by repeatedly calling the simulator:

\begin{Code}
estimator.fit(
    simulator=simulator, epochs=50, num_batches=100, batch_size=64
)
\end{Code}

The arguments \code{epochs}, \code{num_batches}, and \code{batch_size} are exemplarily chosen here and do not necessarily reflect good defaults. 
In online training, there is no finite data set to iterate over: the simulator is called afresh for every batch, so no simulation is ever seen twice. An ``epoch'' is therefore only a helper unit of \code{num_batches} optimizer steps. The product of \code{epochs} and \code{num_batches} determines the total number of gradient steps (and thus the learning-rate schedule). In contrast, the epoch boundary determines how often validation metrics are computed, checkpoints are written, and the loss history is logged.

When we provide simulation data directly to \code{estimator.fit}, this implies offline training. For this purpose, we first define an \code{OfflineDataset} object, which hosts the simulated training data and knows how to split them into batches during each epoch. Then, we pass the data to the \code{dataset} argument of \code{estimator.fit}:

\begin{Code}
sims = simulator.sample(1024)
data = bf.OfflineDataset(sims, batch_size=64, adapter=adapter)
estimator.fit(dataset=data, epochs=50)
\end{Code}

Offline training is relevant for cases where the training data was pre-simulated. It directly stores the simulated data as a dictionary of variables rather than the simulator object. Using offline training is often preferable when the simulator is slow or cannot be easily executed within \proglang{Python}. Moreover, training tends to be faster with offline training even for fast simulators since training data can be pre-loaded into the GPU for more efficient neural network training. However, since the pre-simulated training data is fixed and finite, each simulation will be seen multiple times during neural network training (more precisely: once per epoch). As such, an offline training strategy is susceptible to overfitting. To already see potential overfitting during training, we can pass other simulated data to the \code{validation_data} argument of \code{estimator.fit}:

\begin{Code}
val_sims = simulator.sample(128)
val_data = bf.OfflineDataset(val_sims, batch_size=64, adapter=adapter)
estimator.fit(dataset=data, validation_data=val_data, epochs=50)
\end{Code}

As is customary in deep learning, the loss on the validation data is evaluated and shown in the history but not used to train the neural networks. 
In contrast to offline training, online training is not susceptible to overfitting because each simulation is seen only once during training (hence no need for validation data). 
%We will showcase the practical differences of training strategy choice in our case studies (\autoref{sec:case-studies}).

\subsubsection{Inference with Estimators}
\label{sssec:inference}

Once trained, we can ask the estimator to generate samples from the target distribution or evaluate the target's log-density. For our example, the target is the posterior $p(\mu, \sigma \mid x)$ of parameters $\mu$ and $\sigma$ given data $x$. Since, after training the neural networks, inference is amortized, we can get near instant posterior samples for multiple new (simulated or real-world) datasets at once:

\begin{Code}
test_sims = simulator.sample(10)
samples = estimator.sample(conditions=test_sims, num_samples=100)
\end{Code}

The \code{samples} object is a dictionary with names according to the inference variables (here \code{"mu"} and \code{"sigma"}). 
Each element is a \pkg{numpy} array of shape \code{(n_datasets, n_samples, len_parameter)}. Accordingly, for our example, the samples of both \code{"mu"} and \code{"sigma"} have shape \code{(10, 100, 1)}, since we evaluated 100 samples for each of 10 simulated datasets, and our parameters are both scalar (\code{len_parameter = 1}).

In addition to sampling, we can also evaluate the log-density of the target distribution at given values of both inference variables and conditions. For example:

\begin{Code}
log_densities = estimator.log_prob(data=test_sims)
\end{Code}

The \code{data} argument of \code{estimator.log_prob} needs to contain both the conditions of the target distribution (here $x$) and the values of the inference variables ($\mu$ and $\sigma$) at which the density should be evaluated. This stands in contrast to \code{estimator.sample} whose \code{conditions} argument only requires the conditions of the target but no values of the inference variables themselves, since those are to be sampled.

\subsection{Diagnostics}
\label{ssec:diagnostics}

Just because we have trained an estimator does not mean its approximation to the target distribution is necessarily a good one. In order to validate whether training has been successful, \pkg{BayesFlow} comes with a wide range of diagnostics all hosted within the \code{Diagnostics} module. Within this module, we distinguish between graphical diagnostics (submodule \code{diagnostics.plots}) and numerical diagnostics (submodule \code{diagnostics.metrics}). Usually the first graphical diagnostic investigated after training is the loss history:

\begin{Code}
bf.diagnostics.plots.loss(estimator.history)
\end{Code}

Ideally, it should show the training loss (and validation loss if present) to reduce over the epochs and eventually converge at a low level (see~\autoref{fig:auto-diagnostics} in \autoref{sec:case-studies} for an example). 

Most other graphical diagnostics perform a comparison between the obtained samples from the (approximated) target distribution (argument \code{estimates}) and a simulated ground truth (\code{targets}). For example, to perform graphical simulation-based calibration checking \citep[SBC;][]{modrakSimulationBasedCalibrationChecking2023, talts2020sbc}, we can run:

\begin{Code}
test_sims = simulator.sample(10)
test_samples = estimator.sample(conditions=test_sims, num_samples=100)

bf.diagnostics.plots.calibration_ecdf(
    estimates=test_samples,
    targets=test_sims,
)
\end{Code}

Most numerical diagnostics follow the same pattern. For example, to compute the test statistic corresponding to the above calibration ECDF plot \citep{sailynoja2022sbc}, we can use:

\begin{Code}
bf.diagnostics.metrics.calibration_log_gamma(
    estimates=test_samples,
    targets=test_sims,
)
\end{Code}

More diagnostics and further arguments to tailor them are discussed in \autoref{sec:case-studies}.

\subsection{Workflows}
\label{ssec:workflows}

As an alternative to the lower-level \code{Estimator} module, \pkg{BayesFlow} offers the high-level 
\code{Workflow} interface, from which all steps in the training and inference phase can conveniently be called from.
First, we gather all main objects---the \code{Simulator} (optional), the \code{Adapter (optional)},
and the \code{Networks} (optional summary network)---into one object:

\begin{Code}
workflow = bf.BasicWorkflow(
    simulator=simulator,
    adapter=adapter,
    inference_network=inference_network,
    summary_network=summary_network,
)
\end{Code}

We can then train the workflow with a single line of code, for example, using online training:

\begin{Code}
workflow.fit_online(epochs=50, batch_size=64, num_batches_per_epoch=100)
\end{Code}

Subsequently, we can run many diagnostics at once to obtain a good overview of the obtained approximations:

\begin{Code}
test_data = simulator.sample(10)
workflow.plot_default_diagnostics(test_data=test_data, num_samples=100)
workflow.compute_default_diagnostics(test_data=test_data, num_samples=100)
\end{Code}

To perform actual inference on new data, \code{workflow.sample} and \code{workflow.log_prob} continue 
to work in the same way as for the \code{estimator}. More generally, the \code{workflow} object still 
allows to access all lower level elements and methods of its components, such that its increase
in convenience does not come with a relevant loss in flexibility or expressiveness. In our own applications, we prefer using workflows over their lower level alternatives.

%Use a case study to provide an overview / example using the workflow objects (approx. half a page)
%The high-level model fitting workflow with \pkg{BayesFlow} uses the \code{Workflow} objects of the same name.

\section{Case Study: Lotka-Volterra Dynamics}
\label{sec:case-studies}

To demonstrate \pkg{BayesFlow}'s end-to-end workflow with a more complex scenario than outlined in \autoref{sec:software}, we present a case study in this section using the Lotka-Volterra system \citep{lotka2002contribution, wangersky1978lotka, bunin2017ecological} as observation model. For clarity and ease of understanding, we only highlight the most important code segments and outputs. A fully reproducible tutorial for this case study can be found in the following repository: \href{https://github.com/bayesflow-org/bf2-paper-case-study}{https://github.com/bayesflow-org/bf2-paper-case-study}.

\subsection{Preliminary: Choosing a Backend}

% Before we start using \pkg{BayesFlow}, we need to commit to one of the deep learning backends that it supports. \pkg{TensorFlow}, the default backend for \pkg{Keras 3}, is inherited as the default backend for \pkg{BayesFlow}. However, one can also opt for using \pkg{PyTorch} or \pkg{JAX} instead. In general, the choice of backend depends on several factors, including but not limited to: 1) the available operations that each backend has, 2) the demand for speed and hardware requirements, and 3) compatibility with any external tools included in the workflow.

% Up-To-Date Version:
Before we start using \pkg{BayesFlow}, we need to commit to one of the deep learning backends that it supports. In general, the choice of backend depends on several factors, most importantly, the available hardware and speed requirements, as well as compatibility with already existing code, and external tools used in the workflow.

Once installed, \pkg{BayesFlow} will attempt to automatically detect and use whichever backend the user has installed. This is indicated as follows:

\begin{Code}
import bayesflow as bf
> INFO:bayesflow:Using backend 'torch'
\end{Code}

To use this automatic detection, \pkg{BayesFlow} must be imported before \pkg{Keras}.
We can also override this behavior by setting the \code{KERAS_BACKEND} environment variable before importing \pkg{BayesFlow}, for instance by using the \proglang{Python} built-in \code{os} module.
This is particularly useful when multiple backends are installed in the same environment.

\begin{Code}
import os
os.environ["KERAS_BACKEND"] = "tensorflow"
import bayesflow as bf
> INFO:bayesflow:Using backend 'tensorflow'
\end{Code}

In this case-study, we choose \pkg{JAX} for its fast performance, optimal reproducibility, and functional purity. We set this backend explicitly, and also ensure reproducibility by enabling determinism in XLA and CUDNN:

\begin{Code}
import os

os.environ["XLA_FLAGS"] = "--xla_gpu_deterministic_ops=true --xla_gpu_autotune_level=0"
os.environ["TF_DETERMINISTIC_OPS"] = "1"
os.environ["TF_CUDNN_DETERMINISTIC"] = "1"
\end{Code}

% \pkg{BayesFlow} can work with both CPU and GPU supports for any chosen \pkg{Keras} backend. In general, GPU support has the potential to accelerate training. However, this acceleration is sensitive to simulation budget and model complexity: If the simulation budget is small (or online training is used, which usually runs on the CPU anyway), model complexity is low, and the neural estimators are small, training on CPU might as well be faster than GPU.

\subsection{Observation Model Definition}

As we suggested in \autoref{fig:abi}, defining an observation model involves 1) sampling the governing parameters $\theta$ from their specified prior distributions, 2) sampling some measurement noise $\xi$, and 3) generating observables from an observation model $\mathcal{D} = \operatorname{Sim}(\theta, \xi)$. 

The Lotka-Volterra (LV) system \citep{lotka2002contribution, wangersky1978lotka, bunin2017ecological} presents a classic example in computational ecology that models the predator-prey population dynamics between two species. Formally, the LV system is defined as a set of two nonlinear ODEs:
\begin{align}
    {\frac {\mathrm{d}x}{\mathrm{d}t}}&=\alpha x-\beta xy,\\{\frac {\mathrm{d}y}{\mathrm{d}t}}&=-\gamma y+\delta xy,
\end{align}
where $x$ is the prey population, $y$ is the predator population, $\alpha, \beta$ are the rates of increase and decay for the prey population, while $\gamma, \delta$ are the rates of increase and decay for the predator population, respectively. Together, $\theta = (\alpha, \beta, \gamma, \delta)$ are the governing parameters of the LV system equations.
While we can directly measure the population time series using these equations, we cannot directly measure the governing parameters.

In its original form, we can implement the LV system equations as follows:

\begin{Code}
def lotka_volterra_equations(state, t, alpha, beta, gamma, delta):
    x, y = state
    dxdt = alpha * x - beta * x * y
    dydt = - gamma * y + delta * x * y
    return [dxdt, dydt]    
\end{Code}

However, we still need to make some assumptions about the LV system. First, we need to provide an initial state for the two species. Second, we need to specify a time horizon for the simulation as a whole. Combining these two assumptions, we can create a full implementation of the LV system by solving the ODE using \code{scipy.integrate.odeint}:

\begin{Code}
def ecology_model(
    alpha, beta, gamma, delta, 
    t_span=[0, 5], t_steps=100, initial_state=[1, 1]
):
    t = np.linspace(t_span[0], t_span[1], t_steps)
    state = odeint(lotka_volterra_equations, 
        initial_state, t, args=(alpha, beta, gamma, delta))
    x, y = state.T
    
    return dict(x=x, y=y, t=t)
\end{Code}

Furthermore, we would also assume that the observation model is imperfect. In practice, we can reflect this assumption by 1) introducing measurement noise to the ideal observation \code{x, y}, and \code{t}, and 2) including counting error by choosing a subset of time steps to observe. We introduce these changes after the ideal simulation to ensure realistic simulation outputs.

\begin{Code}
def observation_model(x, y, t, subsample=10, obs_prob=1.0, noise_scale=0.1):
    t_steps = x.shape[0]
    # Add noise to observations and determine the observed time steps
    # ...
    return {
        "observed_x": noisy_x[observed_indices],
        "observed_y": noisy_y[observed_indices],
        "observed_t": t[observed_indices]
    }   
\end{Code}

To complete the definition of our observation model, we sample the priors from a scaled logit-normal distribution:

\begin{Code}
def prior():
    x = np.random.normal(size=4)
    theta = 1 / (1 + np.exp(-x)) * 3.9 + 0.1 
    return dict(alpha=theta[0], beta=theta[1], gamma=theta[2], delta=theta[3])
\end{Code}

We can easily inspect the parameters sampled from the prior and the simulated observables by calling \code{prior()} and \code{observation_model(**ecology_model(**prior()))}. 

\subsection{Simulator}
\label{ssec:lv-simulator}

As mentioned in \autoref{ssec:simulators}, we can use the \code{bf.make_simulator()} function to connect all components of our implementation into a single simulator:

\begin{Code}
simulator = bf.make_simulator([prior, ecology_model, observation_model])
\end{Code}

\begin{figure}[htbp]
    \centering
    \includegraphics[width=\linewidth]{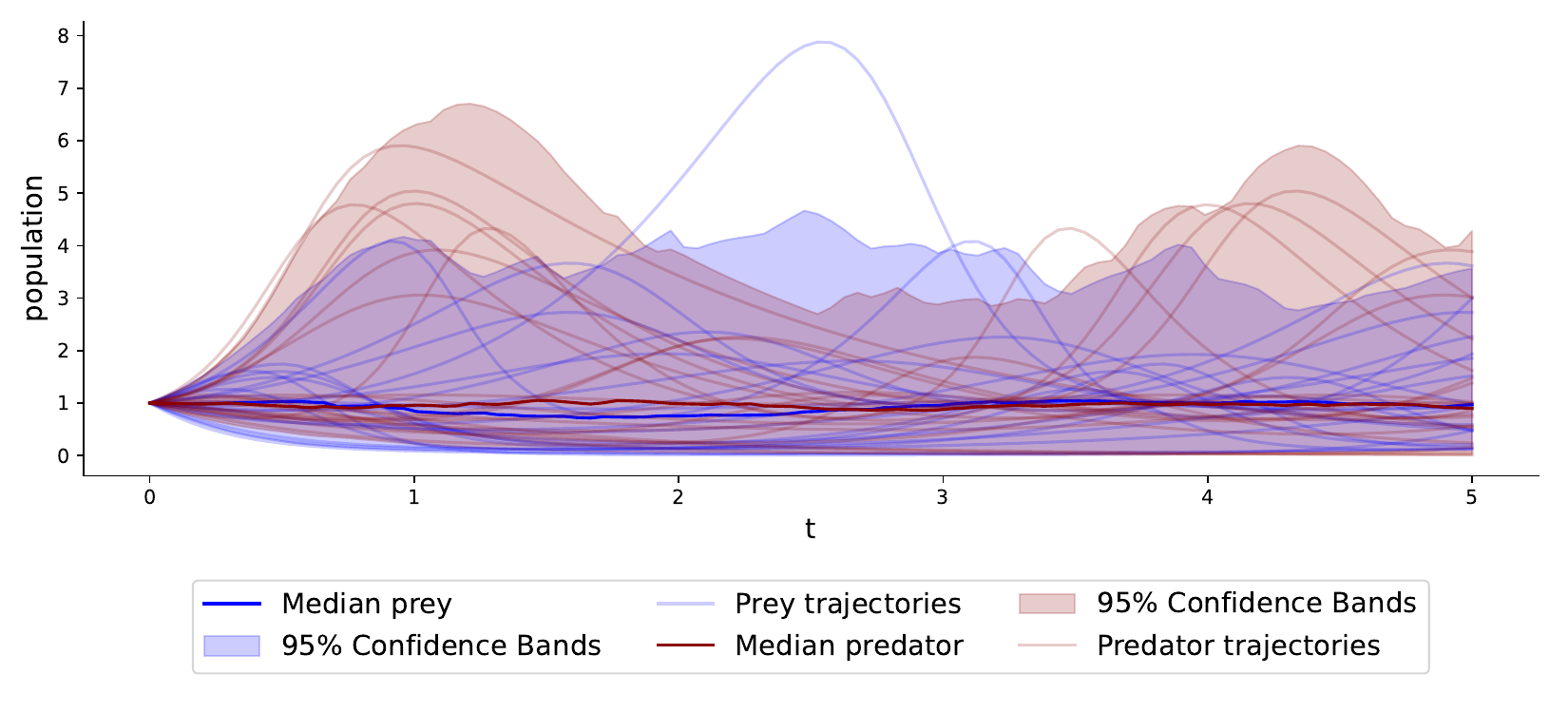}
    \caption{Simulated trajectory aggregates from the LV model with parameters sampled from the prior.}
    \label{fig:sim-trajectories}
\end{figure}

We can then use \code{simulator.sample()} to run multiple simulations. This includes the sets of parameters $\{\theta\}$ as well as the simulated ideal ($\{x, y, t\}$) and observed ($\{x_{\obs}, y_{\obs}, t_{\obs}\}$) time series. 
% We can conveniently use the \code{keras.tree.map_structure()} function to inspect the dimensions of each element in our dataset.
% \begin{Code}
% num_trajectories = 100
% samples = simulator.sample(num_trajectories)
% keras.tree.map_structure(keras.ops.shape, samples)
% {'alpha': (100, 1), 'beta': (100, 1), ... }
% \end{Code}
There are many steps we can take to check our model prior to training, and we recommend following principled prior predictive checks \citep{gelman2020bayesian}. For brevity, we visualize a few the trajectories to observe if the specified priors lead to expected behavior in the observation model. For the LV system, we expect a general oscillating trend for each simulated trajectory pair, differing only by the sampled parameters from the priors (see~\autoref{fig:sim-trajectories}).

% \begin{Code}
% plot_trajectories(samples, ["x", "y"], ["Prey", "Predator"])
% \end{Code}

% Our neural estimator can be used for both online and offline training. When using offline training, a validation dataset is often required for model validation. Since we have already assembled the simulator, we can use the \code{simulator.sample()} function to generate a validation dataset. Here, we choose the simulation budget $N=300$:

% \begin{Code}
% validation_data = simulator.sample(300)
% \end{Code}

\subsection{End-to-End Posterior Estimation With Raw Data}

The most common use of \pkg{BayesFlow} is neural posterior estimation. In this workflow, the dataset generated from the simulator is considered for training the neural estimator, where the parameters' posterior is learned from the simulation outputs. We can easily set up this workflow via the \code{BasicWorkflow} object exactly as outlined in \autoref{fig:workflow}, where we define the \code{simulator}, the \code{adapter}, the \code{summary_network}, and the \code{inference_network}.   
\subsubsection{Creating the Adapter}

As discussed in \autoref{ssec:adapters}, simulation outputs can be ``curated'' into neural network-friendly format via a custom \code{Adapter} object that bundles together multiple stateless transforms:

\begin{Code}
adapter = (
    bf.adapters.Adapter()
    .convert_dtype("float64", "float32")
    .drop(["x", "y", "t"])
    .as_time_series(["observed_x", "observed_y", "observed_t"])
    .concatenate(["alpha", "beta", "gamma", "delta"], 
                into="inference_variables")
    .concatenate(["observed_x", "observed_y", "observed_t"], 
                into="summary_variables")
)
\end{Code}

We can inspect the order of transformations by printing the \code{adapter}. The list below explains the transforms we include here:

\begin{itemize}
    \item The \code{convert_dtype} transform unifies the data type of all simulated data to single-precision floats, the most commonly used precision for deep learning.
    \item The \code{drop} transform removes the unused observables and parameters from the dataset, so that they are not included during training and inference.
    \item The \code{as_time_series} transform ensures that variables are correctly identified as a time-series by corresponding summary networks. Internally, this just adds a trailing dimension of length 1, such that the now second-to-last dimension is the time-series dimension.
    \item the \code{concatenate} transform gathers specific observables or parameters together into a single \pkg{NumPy} array along the last dimension (by default).
\end{itemize}

Concatenation groups the time series as \code{summary_variables} to become the input to the \code{summary_network}; while grouping the model parameters as \code{inference_variables} to become the variables whose posterior distribution is learned by the \code{inference_network}. 

\subsubsection{Defining the Networks}
\label{ssec:e2e-networks}

Next, we can assemble the neural estimator by defining a \code{summary_network} and an \code{inference_network}. Since the observable trajectories are all time series, \pkg{BayesFlow}'s off-the-shelf \code{TimeSeriesNetwork} can be used to compress them into fixed-length summary statistics. The choice of \code{inference_network} is rather free, yet similarly straightforward. Here, we choose the \code{FlowMatching} network for its cost-effectiveness.

\begin{Code}
time_series_network = bf.networks.TimeSeriesNetwork(summary_dim=32)
flow_matching = bf.networks.FlowMatching()
\end{Code}

\subsubsection{Offline Training With the Basic Workflow}

We gather all the ingredients required for training and inference in a \code{BasicWorkflow} object, as detailed in \autoref{ssec:workflows}:

\begin{Code}
workflow = bf.BasicWorkflow(
    simulator=simulator,
    adapter=adapter,
    summary_network=time_series_network,
    inference_network=flow_matching
)    
\end{Code}

Subsequently, we train the neural estimator with an offline dataset consisting of 10k simulations:

\begin{Code}
training_data = workflow.simulate(10000)
history = workflow.fit_offline(
    data=training_data, epochs=80, batch_size=64, 
    validation_data=validation_data
)
\end{Code}

In our case, given our simulation budget, training the neural estimator should be very fast (within the scale of minutes), regardless of whether a GPU is used. 

\subsubsection{Automatic Diagnostics}
After the neural estimator is trained, we perform model validation by probing its computational faithfulness and model sensitivity. The \code{plot_default_diagnostics()} method facilitates this by combining visualizations of convergence, posterior calibration, parameter recovery, and posterior contraction:

\begin{Code}
figures = workflow.plot_default_diagnostics(
    test_data=validation_data,
    variable_names=[r"$\alpha$", r"$\beta$", r"$\gamma$", r"$\delta$"],
    loss_kwargs={...}, recovery_kwargs={...}, calibration_ecdf_kwargs={...},
    z_score_contraction_kwargs={...},
)    
\end{Code}

\begin{figure}[htbp]
    \centering
    \includegraphics[width=\linewidth]{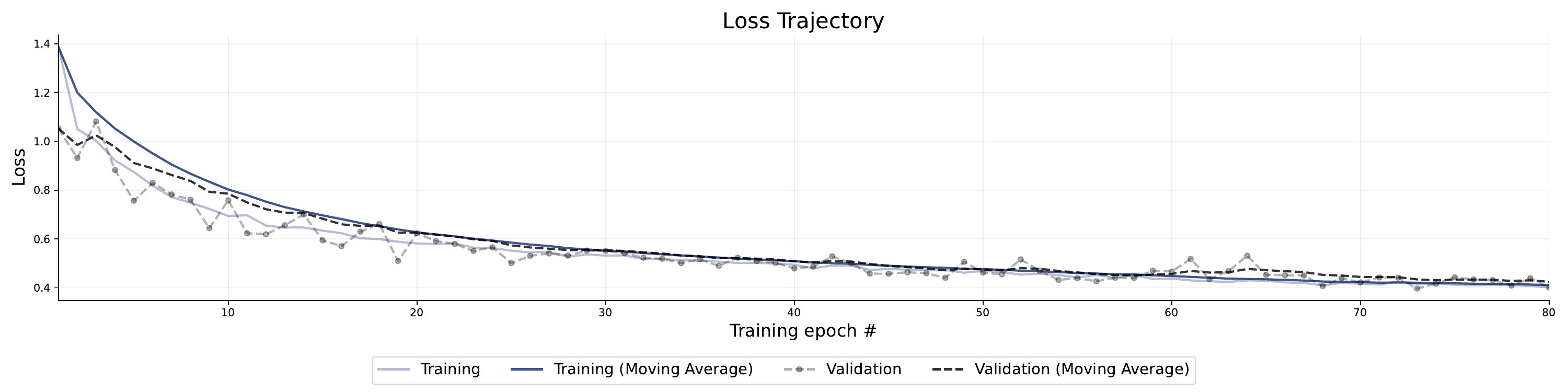}
    \includegraphics[width=\linewidth]{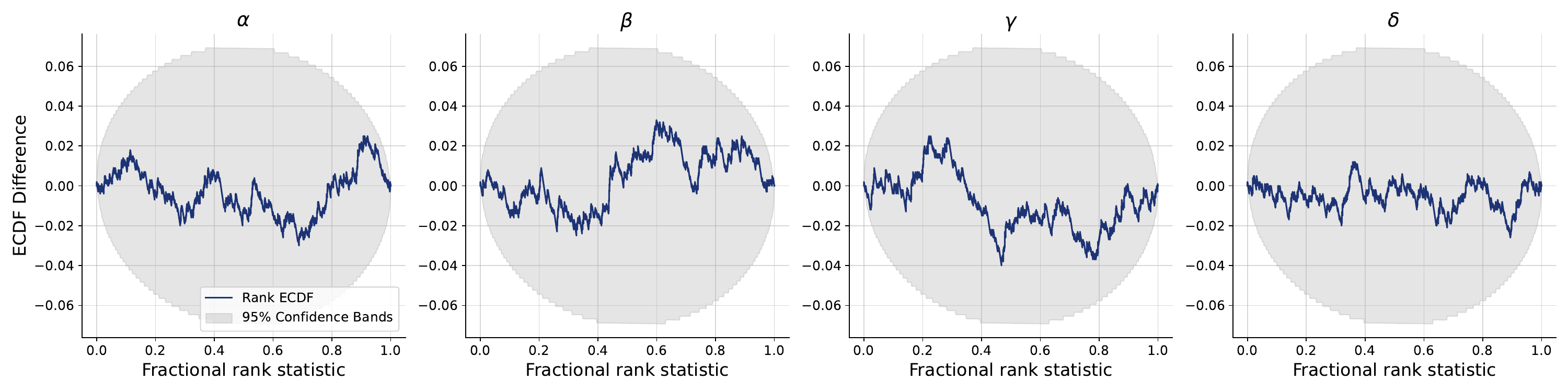}
    \includegraphics[width=\linewidth]{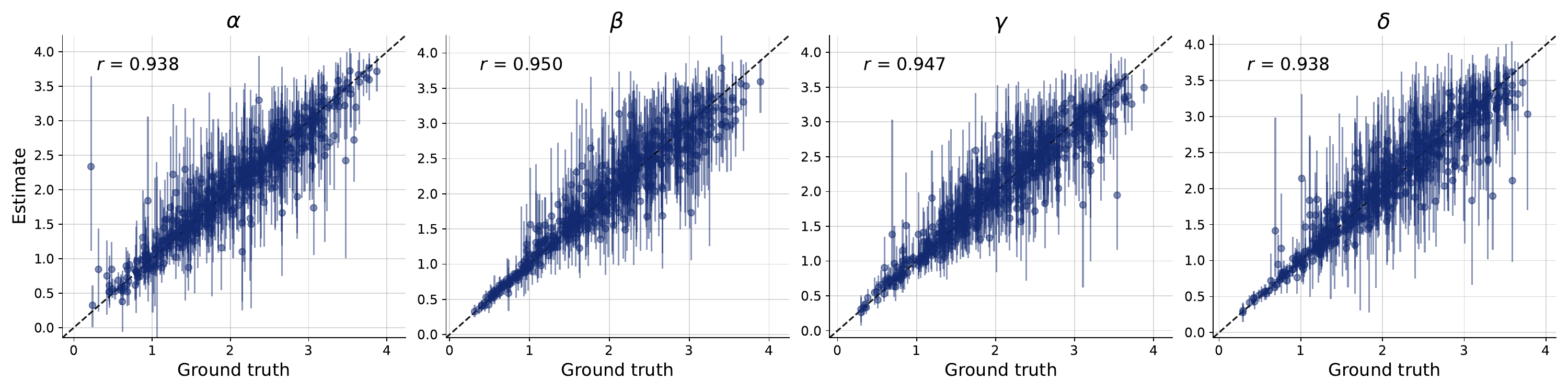}
    \includegraphics[width=\linewidth]{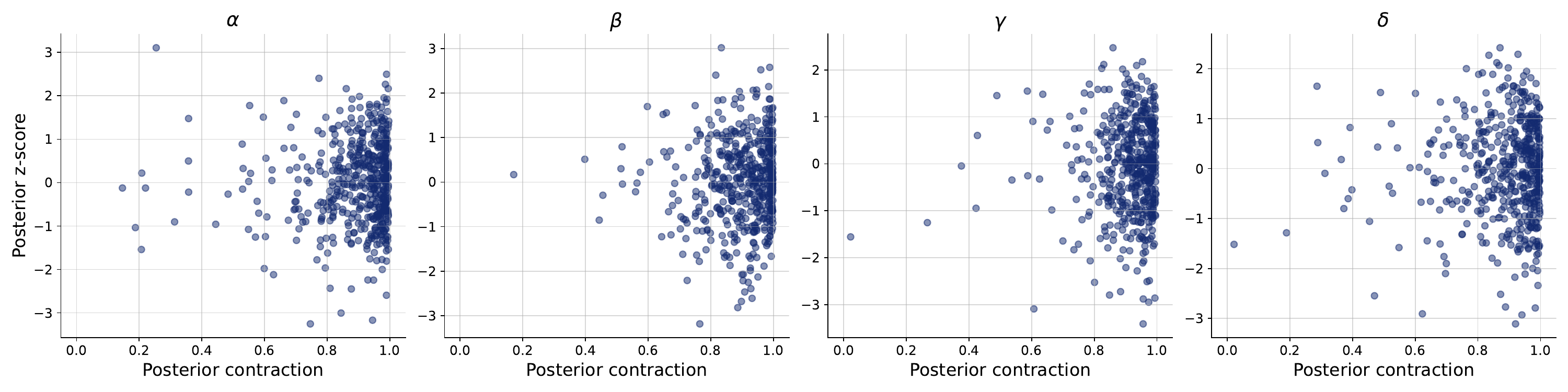}
    \caption{Automatically generated visual diagnostics from the \code{BasicWorkflow} object. \emph{From top to bottom}: training and validation loss, calibration ECDF, parameter recovery, and posterior $z$-score contraction.}
    \label{fig:auto-diagnostics}
\end{figure}

The results (shown in \autoref{fig:auto-diagnostics}) provide a basis for interpreting the reliability of the inference pipeline by monitoring convergence, posterior calibration, and posterior accuracy. Specifically, the loss trajectories indicate that the network has converged without overfitting. The calibration ECDF plots show no evidence of miscalibration: the ECDF differences stay within the confidence bands, so the resulting posterior appears neither over- nor underconfident. Further, the parameter recovery plots indicate good recovery of the ground truth, and the z-score contraction plots suggest the posterior is highly contracted and free of systematic biases. Collectively, these results are consistent with a well-calibrated and informative approximation of the posteriors implied by the LV model, for the given validation data.
%while they cannot rule out errors that are undetectable by these checks \citep{modrakSimulationBasedCalibrationChecking2023}.

The diagnostics shown in \autoref{fig:auto-diagnostics} are summarizing the posteriors obtained for many datasets. We can also investigate the posteriors of individual datasets more closely, for example, via:

%Another informative way to visually inspect how much the neural estimator learned about the parameters from the training data is to plot the posterior distributions over the priors. To this end, \pkg{BayesFlow} provides two visual diagnostics: 1) the $z$-score contraction, which indicates the extent to which posterior means of the parameter differ from the true parameter means, and 2) a direct visualization of posterior distributions over prior distributions given an estimation target (see \autoref{fig:e2e-posterior}). The former is included as part of the default visual diagnostics, while the latter can be visualized by:

\begin{Code}
estimates = workflow.sample(num_samples=500, conditions=validation_data)

f = bf.diagnostics.plots.pairs_posterior(
    estimates=estimates, targets=validation_data, dataset_id=0
)
\end{Code}

\begin{figure}[t]
    \centering
    \includegraphics[width=\linewidth]{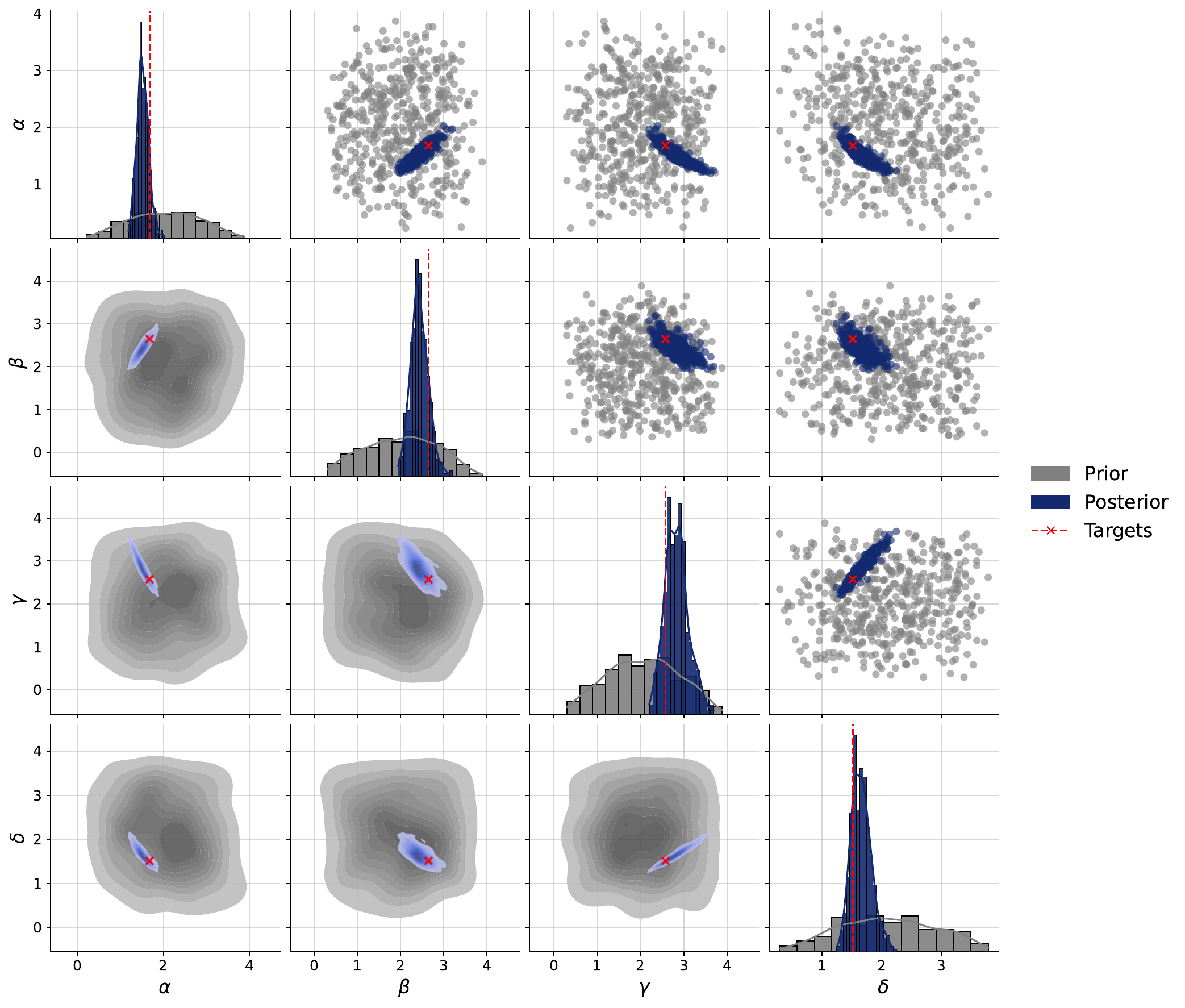}
    \caption{Estimated posterior distribution of the LV model parameters for a single dataset, showing contraction towards the estimation targets compared to the priors.}
    \label{fig:e2e-posterior}
\end{figure}

The \code{dataset_id} argument indicates which sampled dataset (and corresponding posterior) should be displayed in the plot (see \autoref{fig:e2e-posterior}).

In addition to graphical diagnostics, \pkg{BayesFlow} also comes with several numerical diagnostics, such as normalized root mean square error (NRMSE), posterior contraction, and expected calibration error. These diagnostics are readily accessible via the \code{compute_default_diagnostics()} method (see \autoref{tab:e2e-metrics}):

\begin{Code}
workflow.compute_default_diagnostics(test_data=validation_data)
\end{Code}

\begin{table}[t]
\centering
\caption{Diagnostic metrics for posterior estimation. NRMSE is normalized by the prior RMSE. A value of 0 indicates a point-like posterior at the true value, whereas a value of 1 indicates the posterior is as wide as the prior.}
\begin{tabular*}{0.6\textwidth}{@{}lccccc@{}}
\toprule
\textbf{Metrics}        && $\alpha$ & $\beta$ & $\gamma$ & $\delta$ \\
\midrule
NRMSE                   &&0.261     &0.282      &0.260      &0.273   \\
Log Gamma               &&2.267     &4.174      &3.200      &4.612   \\
Calibration Error       &&0.008     &0.021      &0.012      &0.008   \\
Posterior Contraction   &&0.934     &0.925      &0.936      &0.932   \\
\bottomrule
\end{tabular*}
\label{tab:e2e-metrics}
\end{table}

\subsubsection{Posterior Predictive Checks}
\label{sssec:e2e-postcheck}

\begin{figure}[t]
    \centering
    \includegraphics[width=\linewidth]{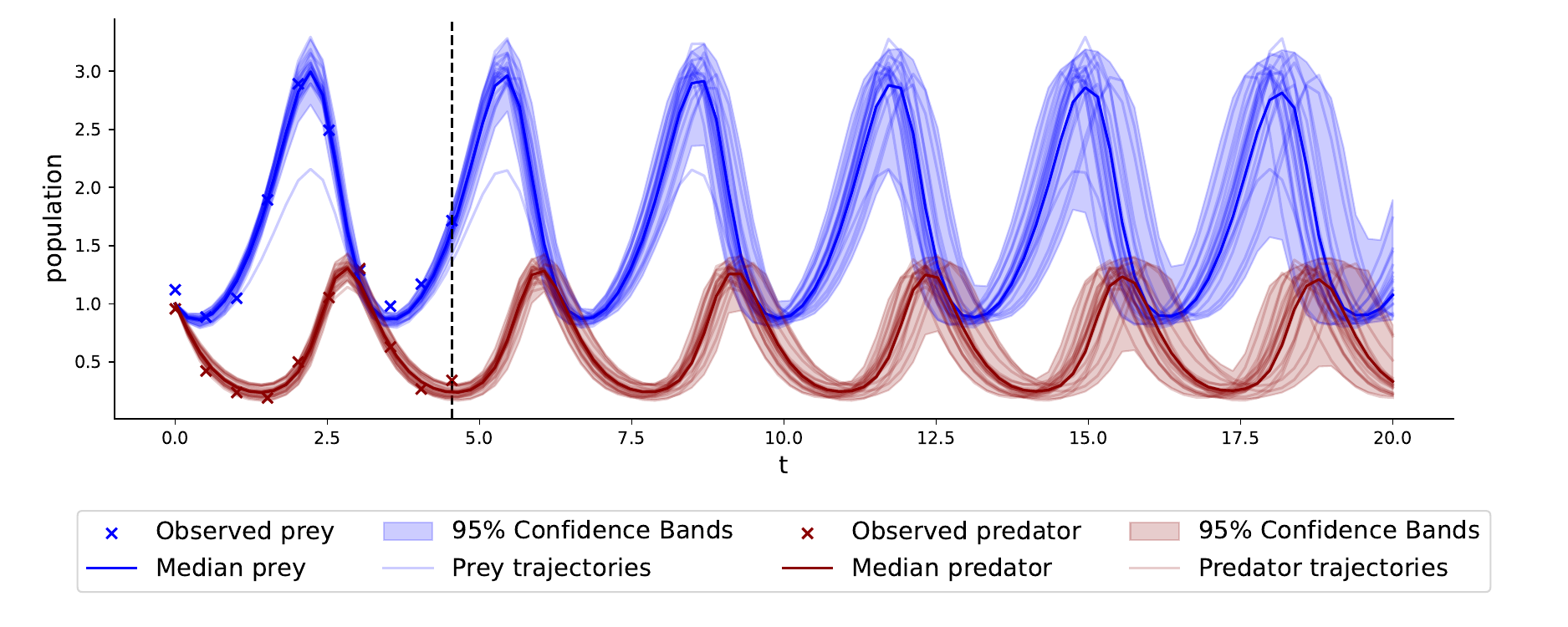}
    \caption{Posterior predictive check of prey-predator population trajectories implied by the LV model for a single validation dataset.}
    \label{fig:e2e-trajectories}
\end{figure}

As a final step, we can use the posterior samples from the neural estimator to resimulate data from the LV model and compare it to the observed data. The goal of such posterior predictive checks is usually to assess how appropriate the statistical model (as encoded by the simulator) is for the given data \citep{gabryVisualizationBayesianWorkflow2019}. In our case, since the validation data has been generated \textit{from} the model, we know the latter is appropriate by definition. Instead, when applied to validation data, we can use posterior predictive checks as another piece of evidence that the neural estimator can indeed approximate the analytic posteriors of the model with sufficient accuracy.

As shown in \autoref{fig:e2e-trajectories}, the posterior-implied trajectories match the observed data well and are much more concentrated than the prior-implied trajectories (compare with \autoref{fig:sim-trajectories}). The model is also capable of predicting trajectories for time horizons going beyond the observed data, with increasing uncertainty as predictions are made further into the future; as would be expected for an appropriately calibrated model and corresponding posterior.

\subsection{Point Estimation with Summary Statistics}
\label{ssec:point-estimation}
In some cases, full posterior estimation may not be feasible, for example, due to large-scale or slow simulator output that limits the simulation budget. In these cases, posterior point estimation becomes a viable alternative to estimating the full posterior \cite{olischläger2026mitigatingrepresentationgapsamortized}. What is more, for the LV model, domain experts have developed summary statistics that can be used to our advantage and accelerate the workflow. Both point estimator and fixed summary statistics allow us to simplify the neural estimator and significantly increase the computational efficiency of our workflow. Here, we demonstrate \pkg{BayesFlow}'s ability to handle these cases.

\subsubsection{Expert Summary Statistics}
\label{sssec:expert-summary}

As shown in \autoref{sssec:e2e-postcheck}, the observed posterior predictive trajectory exhibits characteristics like periodicity, lag, and some degrees of synchronicity (correlations) when compared between predators and prey populations. Therefore, one can extract a collection of summary statistics around these observations with a simple function.

\begin{Code}
def expert_stats(observed_x, observed_y, lags=[2,5]):
    means = np.array([observed_x.mean(), observed_y.mean()])
    ...
    return dict(
        means=means, log_vars=log_vars, 
        auto_corrs=auto_corrs, cross_corr=cross_corr, 
        period=T
    )
\end{Code}

We can include these summary statistics in \pkg{BayesFlow}'s \code{make_simulator()} function.

\begin{Code}
expert_simulator = bf.make_simulator(
   [prior, ecology_model, observation_model, expert_stats]
)
\end{Code}

% Resimulation with \code{expert_simulator} allows us to include these summary statistics as data for the neural estimator.

% \begin{Code}
% samples_with_expert_stats = simulator.sample(3)
% keras.tree.map_structure(keras.ops.shape, samples_with_expert_stats)

% {'alpha': (3, 1), ... 'means': (3, 2), 'log_vars': (3, 2), 
% 'auto_corrs': (3, 4), 'cross_corr': (3, 1), 'period': (3, 1)}
% \end{Code}

\subsubsection{Refining the Adapter}
\label{sssec:point-adapter}

Since we computed the summary statistics already within the simulator, we can directly label them in the \code{Adapter} as \code{inference_conditions}. Naturally, this eliminates the need to use the full trajectories as \code{summary_variables}. 

\begin{Code}
expert_adapter = (
    bf.adapters.Adapter()
    ...
    .concatenate(["means", "log_vars", "auto_corrs", "cross_corr", "period"], 
                  into="inference_conditions")
)
\end{Code}

\subsubsection{Simplifying the Networks}
\label{sssec:point-network}

Since our summary statistics are now precomputed, we no longer need a \code{summary_network} in our workflow. For posterior point estimation, we would need an inference network capable of drawing inference from some strictly proper scoring rules \citep{gneiting2007strictly}. \pkg{BayesFlow}'s \code{ScoringRuleNetwork} allows us to draw inference from mean and quantiles, among others, with two such scoring rules: \code{MeanScore()} and \code{QuantileScore()}:

\begin{Code}
quantile_levels = np.linspace(0.1, 0.9, 5)
scoring_rule_network = bf.networks.ScoringRuleNetwork(
    mean=bf.scoring_rules.MeanScore(),
    quantiles=bf.scoring_rules.QuantileScore(quantile_levels)
)
\end{Code}

We can then create a new \code{BasicWorkflow} that utilizes the \code{scoring_rule_network}:

\begin{Code}
workflow = bf.BasicWorkflow(
    simulator=expert_simulator,
    adapter=expert_adapter,
    inference_network=scoring_rule_network
)
\end{Code}

\subsubsection{Online Training}
\label{sssec:point-training}

Instead of offline training, \pkg{BayesFlow}'s \code{BasicWorkflow} object also supports online training via the \code{fit_online()} method. Instead of computing the training steps using the simulation budget for the training dataset, online training requires the user to specify the number of batches per epoch as the number of training steps. As explained in \autoref{sssec:training}, each of the 50 epochs below consists of 200 freshly simulated batches; no simulation is reused. 

\begin{Code}
history = workflow.fit_online(
    epochs=50, num_batches_per_epoch=200, batch_size=64,
)
\end{Code}

One of the benefits of online training is that, since training data is provided on-the-fly via simulation, it minimizes the possibility of overfitting to the pre-generated training dataset. However, the tradeoff is time. Compared to offline training, where simulation time is limited by the size of the training and validation datasets, online training requires training data to be generated per step, which adds to the total training time.

\begin{figure}[t!]
    \centering
    \includegraphics[width=\linewidth]{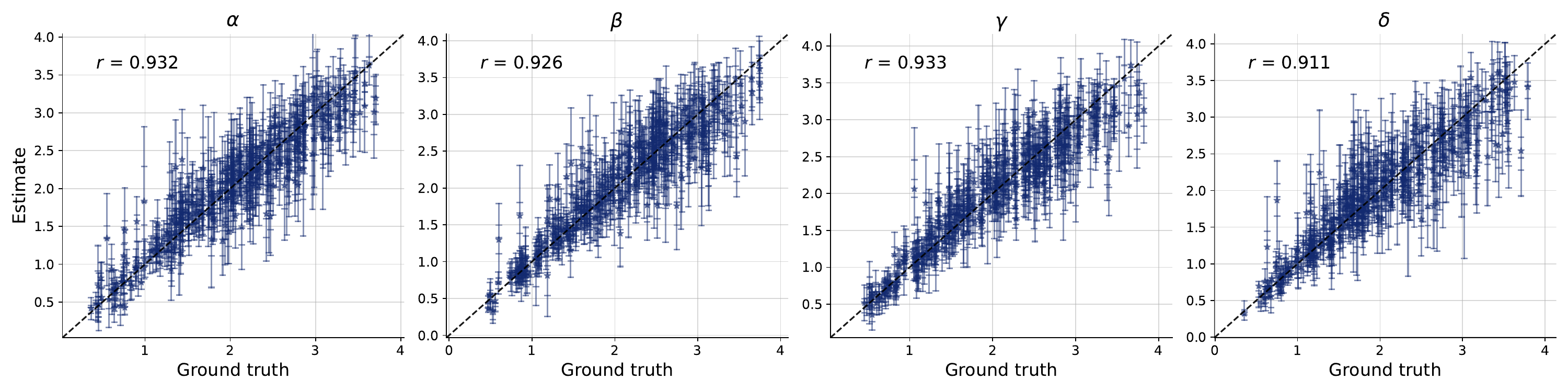}
    \includegraphics[width=\linewidth]{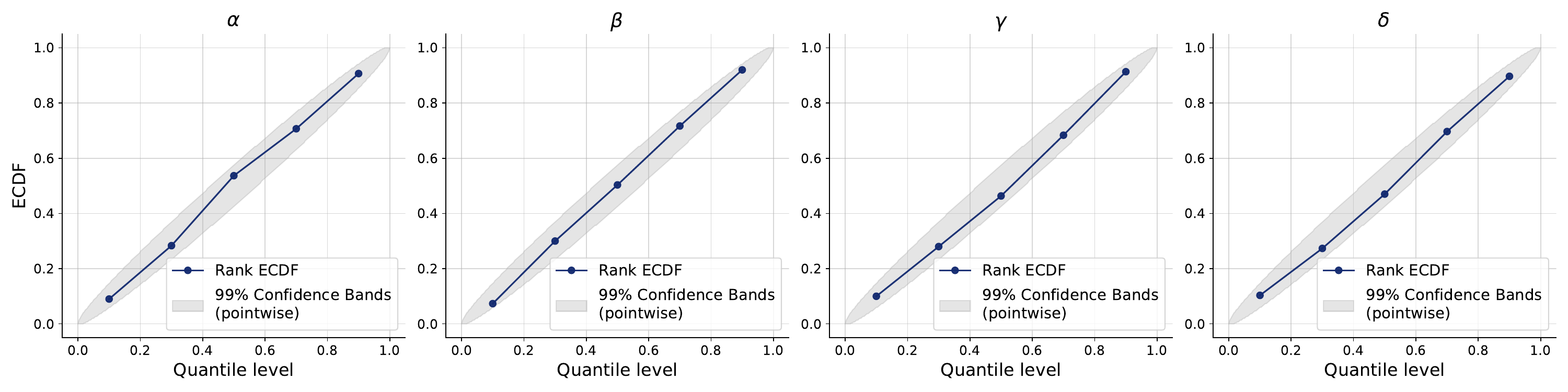}
    \caption{Manual diagnostics for the point estimation, including parameter recovery (\textit{top}) and calibration ECDF (\textit{bottom}).}
    \label{fig:point-diagnostics}
\end{figure}

\subsubsection{Manual Diagnostics}
\label{sssec:manual-diagnostics}

Since this workflow involves point estimation with customized quantile statistics, automatic diagnostics do not apply. Nevertheless, \pkg{BayesFlow} has a diagnostic toolkit available for point estimates, which we can manually extend via customized metrics.
As before, the loss function can be inspected via \code{bf.diagnostics.loss()}. We can obtain estimates of posterior mean and quantiles for all model parameters.

\begin{Code}
validation_data = expert_simulator.sample(500)

point_estimates = workflow.estimate(conditions=validation_data)
keras.tree.map_structure(keras.ops.shape, point_estimates)

{'alpha': {'mean': (500, 1), 'quantiles': (500, 5, 1)}, 
 'beta': ...}
\end{Code}

From here, we can apply visual diagnostics that indicate the neural estimator's accuracy and calibration just using the posterior mean and quantile estimates:

\begin{Code}
f = bf.diagnostics.recovery_from_estimates(
    estimates=point_estimates, 
    targets=validation_data, 
    variable_names=variable_names,
    marker_mapping=dict(quantiles="_", mean="*")
)
\end{Code}

\begin{Code}
f = bf.diagnostics.calibration_ecdf_from_quantiles(
    estimates=point_estimates, 
    targets=validation_data,
    quantile_levels=quantile_levels,
    difference=False,
    variable_names=variable_names
)
\end{Code}

At first glance, visual diagnostics are similar to those for the full posterior estimation (see \autoref{fig:point-diagnostics}). However, the key difference here is that the diagnostics can only show what information is provided by the point estimates. The most evident example of this is the calibration diagnostic, which is only evaluated at the five estimated quantiles. Another difference, not visible in the above plots, is that the per-parameter point estimates do not contain any information on the dependency between parameters. Said dependencies would be required, for example, to obtain complete posterior pairs plots.

%\subsubsection{Posterior predictive check}

%\begin{figure}[htbp]
%    \centering
%    \includegraphics[width=\linewidth]{Figures/point_trajectories.png}
%    \caption{Posterior sample distribution for the LV model parameters showing contraction towards the estimation targets compared to the priors.}
%    \label{fig:point-trajectories}
%\end{figure}

%We can resimulate the posterior trajectories using the neural point estimator. As shown in \autoref{fig:point-trajectories}, we observe that the trajectories obtained from the point estimates exhibit more uncertain patterns of periodicity and synchronicity in the predator-prey dynamics as we extend the time horizons compared to the full posterior estimator (\autoref{fig:e2e-trajectories}). This makes sense given that we leveraged expert summary statistics without using a summary network for compression and estimated only posterior point estimates. 

\begin{table}[t!]
\centering
\caption{Feature comparison of amortized methods in simulation-based inference packages. Entries refer to dedicated functionality exposed by each package's API.}
\begin{tabular}{lccc}
\toprule
 & \pkg{BayesFlow} & \pkg{sbi} & \pkg{NeuralEstimators.jl} \\
\midrule
\textbf{Supported Deep Learning Frameworks} & & & \\
\py \, TensorFlow & \yes & \no  & \no \\
\py \, PyTorch    & \yes & \yes & \no \\
\py \, JAX        & \yes & \no  & \no \\
\jl \, Flux       & \no  & \no  & \yes \\

\addlinespace
\textbf{Supported Model Types} & & \\
Sequential (flat) models & \yes & \yes & \yes \\
Multilevel models           &  \yes  & \yesnt$^{a}$ & \no \\
Multimodal models  & \yes &  \yesnt  & \yesnt \\
Mixed-type models  & \yes  & \yes  & \no \\

\addlinespace
\textbf{Supported Neural Estimators} & & \\
Posterior estimation & \yes & \yes & \yes \\
Likelihood estimation & \yes$^{b}$ & \yes & \yes \\
Ratio estimation & \yes$^{b}$ & \yes & \yes \\
Scoring rule estimation & \yes & \no & \yes \\
% only NR.jl has interval score

\addlinespace
\textbf{Supported Density Estimators} & & \\
Mixture of Gaussians & \yes & \yes & \yes \\
Normalizing flows & \yes & \yes & \yes \\
Diffusion models    & \yes & \yesnt & \no \\
Flow matching         & \yes & \yesnt & \no \\
Free-form flows       & \yes & \no & \no \\
Consistency models     & \yes & \no & \no \\
Ensembles & \yes & \yes & \yes \\
% only NR.jl has mixture of gaussians
% bayesflow has currently only single gaussian

% \addlinespace
% \textbf{Supported Information Type} & & & \\
% Simulation-based & \yes & \yes &  \\
% Likelihood-based & \yes & \yes &  \\
% Mixed & \yes & \yes &  \\

\addlinespace
\textbf{Supported Bayesian Tasks} & & \\
Parameter estimation & \yes & \yes & \yes \\
Model comparison & \yes & \no & \no \\
Model misspecification detection & \yes & \yes & \no \\

\bottomrule
\end{tabular}
\begin{tabularx}{\linewidth} { 
  >{\raggedright\arraybackslash}X 
  >{\centering\arraybackslash}X 
  >{\raggedleft\arraybackslash}X}
\yes \color{black} \, Supported & \yesnt \color{black} \, Partially Supported & \no \color{black} \, Not Supported
\end{tabularx} \\
{\footnotesize $^{a}$ Non-amortized, via integration with \pkg{pyro}. $^{b}$ Posterior sampling via \pkg{PyMC} requires \pkg{JAX}.}
\label{tab:software-comparison}
\end{table}

\section{Comparison Between Packages}
\label{sec:comparison}

Several software packages facilitate amortized Bayesian inference.
A comprehensive comparison of all existing approaches would be beyond the scope of this paper and of limited practical value.
We therefore restrict attention to a small set of direct competitors that are actively maintained, widely used, and provide state-of-the-art inference methods. An overview of the comparisons for amortized methods can be found in \autoref{tab:software-comparison}.

In particular, we focus on \pkg{sbi} \citep{boelts2024sbi} and \pkg{NeuralEstimators.jl} \citep{sainsbury2024likelihood}, which are conceptually closest to \pkg{BayesFlow} in that they support related classes of neural estimators, including posterior, likelihood, and ratio estimators. Other packages, such as \pkg{sbijax} \citep{dirmeier2024sbijax}, \pkg{lampe} \citep{rozet2021lampe}, \pkg{pydelfi} \citep{Alsing2019pydelfi} or \pkg{carl} \citep{Cranmer2015carl}, either target a more limited subset of SBI methods, are at an early stage of development, or have not seen sustained development in recent years. Further tools, including \pkg{madminer} \citep{brehmer2020madminer} and \pkg{swyft} \citep{Miller2022swyft}, are more specialized and emphasize specific estimator types or application domains rather than offering a general-purpose framework for SBI.

\pkg{BayesFlow} is unique among these frameworks in that it natively supports multiple deep-learning backends. This capability arises from a backend-agnostic implementation on top of \pkg{Keras~3}, which allows users to run models using \pkg{PyTorch}, \pkg{JAX}, or \pkg{TensorFlow} backends. By contrast, \pkg{sbi} is based on \pkg{PyTorch}, which may be a friction point for users committed to \pkg{TensorFlow} or \pkg{JAX} ecosystems. Moreover, effective compilation of \pkg{PyTorch} models is still an ongoing effort and can incur overhead due to graph breaks or compatibility issues from lack of operation support. As such, in practice, large \pkg{PyTorch} models often remain uncompiled, resulting in significant performance gaps, especially compared to models written in \pkg{JAX}. Differently, the \pkg{NeuralEstimators.jl} library targets the \proglang{Julia} ecosystem using \pkg{Flux} as the deep-learning backend.

Like \pkg{BayesFlow}, \pkg{sbi} and \pkg{NeuralEstimators.jl} support flat (i.e., single-level) statistical models. While \pkg{sbi} supports non-amortized inference of multilevel models through external integration with \pkg{pyro}, \pkg{BayesFlow} natively accommodates training with multilevel models with arbitrarily nested and crossed probabilistic factorizations as well as varying numbers of parameters. This enables end-to-end amortization of multilevel models within a single framework.

All three packages implement the core SBI estimators --- neural posterior, neural likelihood, and neural ratio estimation. \pkg{BayesFlow} additionally implements a workflow for model comparison with various classification losses \citep{radev2021evidential, jeffrey2024evidence}.
\pkg{BayesFlow} and \pkg{NeuralEstimators.jl} also offer point estimators of posterior and likelihood, enabling rapid prototyping where a point estimate is sufficient before full generative uncertainty quantification is required.

Normalizing Flows are also supported by all three packages: \pkg{sbi} provides masked autoregressive flows \citep{NIPS2017_6c1da886}, neural spline flows \citep{NEURIPS2019_7ac71d43} and allows using flows implemented in \pkg{zuko} \citep{zuko} and \pkg{nflows} \citep{nflows}. BayesFlow ships with affine coupling flows \citep{realnvp} and neural spline flows, whereas \pkg{NeuralEstimators.jl} provides only affine coupling flows.
Additionally, \pkg{BayesFlow} allows for flexible latent spaces, for instance, Student-t for enhanced robustness or mixture distributions for enhanced expressivity \citep{papamakarios2021normalizing}.

Furthermore, \pkg{BayesFlow} incorporates a wide range of modern free-form generative networks, including various flavors of flow matching (e.g., optimal transport, conditional optimal transport), diffusion models (compatible with any schedule, guidance, and composition), and many other recent model families \citep{arruda2025diffusion}, such as few-step (discrete and continuous) consistency models \citep{schmitt2024consistency} and free-form flows \citep{draxler2024freeformflowsmakearchitecture}.

Crucially, \pkg{BayesFlow} offers a unified treatment of generative network architectures, enabling drop-in support for all of its generative architectures for both posterior and likelihood estimation. In contrast, \pkg{sbi} supports diffusion models and flow matching only for posterior estimation via separate, specialized classes with limited customization and \pkg{NeuralEstimators.jl} does not provide generative free-form architectures.

Unlike \pkg{BayesFlow}, \pkg{sbi} offers native support for non-amortized methods such as sequential and multi-round posterior, likelihood, and ratio estimation, as well as several non-amortized sampling methods like MCMC via \pkg{pyro} \citep{bingham2019pyro}, importance sampling, rejection sampling, or variational inference. \pkg{BayesFlow} instead focuses mostly on amortized methods, encouraging direct sampling from a trained amortized estimator or plugging pre-trained NLE/NRE networks into gold-standard MCMC samplers provided by \pkg{PyMC} \citep{abril_pymc_2023} or \pkg{blackjax} \citep{cabezas2024blackjax}.

While all three packages allow bundling pre-trained networks into a single ensemble estimator, only \pkg{BayesFlow} enables bundling \textit{untrained} networks for joint training and allows for arbitrary sharing of summary or inference networks among ensemble members. 
Treating the ensemble as a unified composite network also maximizes hardware utilization, replacing sequential iteration over members with a single, joint training step.
\pkg{NeuralEstimators.jl} supports joint training of point estimators and allows to bundle multiple untrained summary networks\footnote{Using the Flux built-in \texttt{Parallel} layer.}, but other attractive settings like training multiple inference networks using a shared summary network are out of reach in both \pkg{NeuralEstimators.jl} and \pkg{sbi}.

Offline training with pre-computed simulations stored in CPU memory covers many use cases already and is supported in all packages.
To address demand for large-scale scientific problems, \pkg{BayesFlow} further allows users to choose from an array of training schemes, most notably online training, where the simulator is called on-the-fly, generating new samples as the training progresses, as well as offline training using samples stored on-disk instead of in-memory.
Similarly, \pkg{NeuralEstimators.jl} allows users to pass a sampler as a dataset, enabling on-the-fly training, but leaves it to users to implement their own custom on-disk dataset.

Finally, all packages provide diagnostics to assess the performance of trained neural estimators.
\pkg{sbi} and \pkg{BayesFlow} both provide a comprehensive suite of utilities, metrics, and plots to carry out recovery and simulation-based calibration (SBC) checks, including optional custom test quantities \citep{modrakSimulationBasedCalibrationChecking2023}, as well as model misspecification detection \citep{schmitt2023detecting} and sample-based comparison with reference distributions, like C2ST \citep{lopez-paz2018} and MMD \citep{gretton2012}.
\pkg{NeuralEstimators.jl} primarily focuses on assessing the Bayes risk and frequentist coverage of estimates and  ships with convenient recovery plots.

\section{Conclusion}

In this work, we presented \pkg{BayesFlow} Version 2, a backend-agnostic and feature-rich framework for scalable ABI workflows. We demonstrate \pkg{BayesFlow}'s intuitive and comprehensive API by outlining the software's core components and applying the framework to a representative case study. By natively integrating recent state-of-the-art deep learning architectures such as various flavors of diffusion models and flow matching, \pkg{BayesFlow} enables rapid, flexible, end-to-end Bayesian inference across a broad range of tasks. We further showcased \pkg{BayesFlow}'s \code{diagnostics} module, which enables comprehensive inspection of trained estimators to ensure estimation quality through numerical and visual diagnostics.

% We position \pkg{BayesFlow} as the gold-standard toolkit of choice for SBI. Through our demonstration of its software components and associated workflows, we have shown its comprehensiveness of features, as well as the ease of understanding of its interface. With state-of-the-art deep learning architectures as the building blocks of neural estimator, we can use \pkg{BayesFlow} to reliably perform a variety of Bayesian tasks. 

\pkg{BayesFlow} provides a streamlined interface that closely reflects established ABI terminology, carefully aligning naming conventions from statistics and ABI with those used in modern machine learning. The roles of the simulator, neural estimator, and diagnostic utilities are clearly delineated, providing users with a well-structured workflow for generative model specification, data pre-processing, training, and inference. By reducing conceptual overhead and lowering the entry barrier, we argue that \pkg{BayesFlow}'s new design extends its appeal to a broader community of scientists who rely on model-based inferences in their research. Ultimately, the streamlined interface establishes a clear conceptual foundation that supports accessibility for new users and preserves flexibility for advanced practitioners.

% \pkg{BayesFlow}'s streamlined interface is tightly coupled with the semantics of SBI terminology. In particular, the role of simulator, neural estimator, and diagnostics utilities are clearly partitioned and distinguished within the workflow, where a clear structure is presented to the user about data generative generative model specification, data pre-processing, training, and inference. As such, the interface logic is also closely aligned with modern ML workflows. We argue that this streamlined interface allows us to reach a broader user base where simulation is essential to their scientific workflow. (Add a sentence here proposing how.) (Add a concluding sentence to elevate the significance of the streamlined interface).

For the future, we have several plans on how to further improve the functionality of \pkg{BayesFlow}. 
The field of generative modeling is quickly changing, with new architectures and architecture improvements being developed constantly. 
We will continue to implement them in \pkg{BayesFlow} as they become available and reveal promising results for ABI.
This will also include modular architecture combinations that mirror probabilistic symmetries of the approximated statistical
model \citep{habermann_amortized_2024, habermann2026multilevel, kucharsky2025amortized, elsemuller2024deep} as well as loss functions specific to ABI that increase its accuracy and robustness \citep{schmitt_leveraging_2024, mishra2025robust, elsemueller2025does}. 

Moreover, we are currently building \pkg{BayesFlow} interfaces with other probabilistic programming languages, such as \proglang{Stan} \citep{stan_2025}. This would strongly increase the potential user base of \pkg{BayesFlow} and overall simplify specification of simulators and corresponding log densities.
Finally, we will continue to implement modern training methods that work out-of-the-box, such as automatic low-precision pre-training, to support efficient and flexible scaling to the largest Bayesian problems.

\section*{Acknowledgments}
This work was partially funded by the National Science Foundation under Grant No.~2448380, the Deutsche Forschungsgemeinschaft (DFG, German Research Foundation) Projects 528702768 and 508399956 as well as DFG Collaborative Research Center 391 (Spatio-Temporal Statistics for the Transition of Energy and Transport) -- 520388526.
\begin{leftbar}
%All acknowledgments (note the AE spelling) should be collected in this
%unnumbered section before the references. It may contain the usual information
%about funding and feedback from colleagues/reviewers/etc. Furthermore,
%information such as relative contributions of the authors may be added here (if any).
\end{leftbar}

%% -- Bibliography -------------------------------------------------------------
%% - References need to be provided in a .bib BibTeX database.
%% - All references should be made with \cite, \citet, \citep, \citealp etc.
%%   (and never hard-coded). See the FAQ for details.
%% - JSS-specific markup (\proglang, \pkg, \code) should be used in the .bib.
%% - Titles in the .bib should be in title case.
%% - DOIs should be included where available.

\newpage
\bibliography{refs}

%% -- Appendix (if any) --------------------------------------------------------
%% - After the bibliography with page break.
%% - With proper section titles and _not_ just "Appendix".

\newpage

\begin{appendix}

\end{appendix}

%% -----------------------------------------------------------------------------

\end{document}